\documentclass[usenatbib]{mnras}

\usepackage{mathptmx}
\usepackage{amssymb}
\usepackage[T1]{fontenc}
\usepackage{flushend}
\usepackage{stfloats}
\usepackage{float}
\DeclareRobustCommand{\VAN}[3]{#2}
\let\VANthebibliography\thebibliography
\def\thebibliography{\DeclareRobustCommand{\VAN}[3]{##3}\VANthebibliography}

\usepackage{graphicx}
\usepackage{amsmath}
\usepackage{subcaption}
\usepackage{makecell}

\graphicspath{{figure/}}

\makeatletter
\newenvironment{inplacetable}{%
  \let\@makecaption\SFB@maketablecaption
  \@float{table}[H]\fstyle@table}{\endtable}
\makeatother

\makeatletter
\renewcommand{\SFB@maketablecaption}[2]{%
  \vskip 6pt
  {\reset@font\small\centering{\bf #1}%
    \if@cont \ -- {\it continued}\else{\bf .}\fi\ {\rm #2}\par}%
  \vskip 6pt
  \@contfalse
}
\makeatother

\title[Fermi-LAT Detection of SNR G156.2+05.7]{Fermi-LAT Detection of Extended GeV Emission toward the Middle-aged Supernova Remnant G156.2+05.7}

\author[W.-L. Xu et al.]{
Wei-Liang Xu,$^{1}$
Qing-Wen Tang,$^{1}$\thanks{E-mail: qwtang@ncu.edu.cn}
Long Li,$^{1}$
\\
$^{1}$Department of Physics, School of Physics and Materials Science, Nanchang University, Nanchang 330031, China
}

\date{Accepted 2026 November 2. Received 2026 November 2; in original form 2026 June 10}
\pubyear{2026}

\begin{document}
\label{firstpage}
\pagerange{\pageref{firstpage}--\pageref{lastpage}}
\maketitle

\begin{abstract}
We analyse 17 years of Fermi-LAT $\gamma$-ray data toward the middle-aged supernova remnant (SNR) G156.2+05.7, 
supplemented by radio and CO observations and by published X-ray constraints. An extended $\gamma$-ray source is detected, 
well described by a 2D Gaussian template ($R_{68} \sim 0.84^\circ$). This description is preferred over a foreground CO template, whereas the Gaussian and a uniform disc are statistically indistinguishable under the AIC. The spectrum shows evidence for curvature, with a log-parabola preferred over a power law at the $\approx 3.4\sigma$ level. Multi-wavelength spectral energy distribution (SED) modelling, under the adopted assumptions, suggests a leptonic--hadronic joint scenario consistent with the data, with the Fermi-LAT spectrum favouring a hadronic contribution over a purely leptonic interpretation. However, the proton-dominated alternative remains competitive, so the partition of the GeV emission is model-dependent. In the adopted leptonic--hadronic model, the electron population accounts for the radio synchrotron emission and a leptonic $\gamma$-ray component, while a steep proton population contributes the remaining hadronic $\gamma$-rays. Adopting a reference distance $d = 2.7$ kpc, the fitted electron cutoff ($E_{e,\mathrm{cut}}\approx 1.2$ TeV) is compatible with the thermal-dominated X-ray emission, while the proton cutoff is fixed by assumption at $1$ TeV. The northeast and southwest quadrants 
show enhanced $\gamma$-ray emission, which may be related to a clumpy or asymmetric surrounding medium. A correspondence with the directions of optical H$\alpha$ filaments is suggested by the quadrant orientation but is not yet established. No significant long-term flux variability is found, consistent with steady emission from an extended source.
\end{abstract}


%%%%%%%%%%%%%%%%%%%%%%%%%%%%%%%%%%%%%%%%%%%%%%%%%%

%%%%%%%%%%%%%%%%% BODY OF PAPER %%%%%%%%%%%%%%%%%%

\section{Introduction}
\label{sec:intro}

Galactic cosmic rays (CRs) up to the knee are widely thought to be accelerated at supernova remnant (SNR) shocks 
\citep{bell1978acceleration,blandford1987particle,blasi2013origin}. The $\sim10^{51}\,\mathrm{erg}$ released in a core-collapse 
or thermonuclear explosion drives blast waves into the ambient interstellar medium (ISM), producing filamentary shocked 
plasma and broadband non-thermal emission from radio to $\gamma$-rays. Characterising this $\gamma$-ray emission is 
therefore central to tracing the acceleration sites and resolving the century-old puzzle of CR origin \citep{funk2015ground}.

SNR $\gamma$-rays can arise through leptonic channels---inverse Compton (IC) scattering and bremsstrahlung of 
relativistic electrons, with synchrotron dominating the radio-to-X-ray range---or through hadronic channels \citep{reynolds2008supernova,drury1994gammaray}, where 
accelerated protons collide with ambient gas to produce neutral pions that decay via $\pi^{0}\rightarrow2\gamma$. 
The pion-decay signature, detected in several SNRs \citep{ackermann2013detection}, provides a direct 
probe of relativistic nuclei. Fermi-LAT observations therefore serve two purposes. They discriminate between
leptonic and hadronic emission scenarios, and they assess whether the observed morphology traces the
remnant, foreground gas, or localised density structure.

Diffusive shock acceleration (DSA) at SNR blast waves converts a fraction of the kinetic energy into 
non-thermal particles \citep{bell1978acceleration,blandford1987particle}. At the highest energies, 
scattering efficiency drops as the diffusion length grows, allowing some accelerated CRs to escape upstream. 
These particles can produce $\gamma$-ray emission from nearby atomic or molecular gas and give rise to extended GeV--TeV halos whose spectra 
encode the escaping CR population and the local diffusion properties 
\citep{aharonian1996emissivity,gabici2009broadband,li2010grays}.

Middle-aged remnants, by contrast, often evolve in strongly non-uniform
environments. As the shocks slow, they can become radiative and interact with dense
clumps, and cloud crushing and turbulence can locally enhance the target density and
magnetic field even when the volume-averaged interior remains tenuous
\citep{fesen1980spectrophotometry,hartigan1987radiative,bell2004turbulent}. Such
localised structures are distinct from the extended CR halos described above.

G156.2+05.7, discovered as an X-ray shell by \emph{ROSAT} \citep{pfeffermann1991discovery}, stands out 
for its bright thermal X-rays and unusually faint radio shell \citep{reich1992radio}. For decades, its 
distance and age remained unsettled, complicating any evolutionary interpretation. Early Sedov modelling 
favoured a middle-aged remnant at $\sim1$--$3\,\mathrm{kpc}$ and an age of several $\times10^{4}\,\mathrm{yr}$ 
\citep{pannuti2004asca}, while 6-cm polarisation maps revealed no central compact object 
and suggested contamination by foreground structures \citep{xu2007polarization}. Optical H$\alpha$ imaging 
exposed filamentary emission aligned with dust lanes and interstellar clouds, prompting some authors to 
advocate a closer, younger picture \citep{gerardy2007discovery}. Multi-point \emph{Suzaku} spectroscopy 
then showed that the hard X-ray component, after subtraction of the background cluster 
2XMM~J045637.2+522411, is consistent with the cosmic X-ray background, leaving no measurable non-thermal 
X-ray excess---consistent with an evolved remnant whose shocks have slowed \citep{uchida2012suzaku}.

A key constraint came from proper-motion measurements of nonradiative H$\alpha$ filaments at the 
rim. \citet{katsuda2016constraining} combined the tangential expansion with X-ray-derived shock velocities 
to place a lower limit $d \gtrsim 1.7\,\mathrm{kpc}$ and an age $\gtrsim 2\times10^{4}\,\mathrm{yr}$, 
supporting a middle-aged interpretation for G156.2+05.7. For definiteness, we adopt $d = 2.7$ kpc as the reference distance for all distance-dependent quantities, consistent with this proper-motion lower limit and the earlier Sedov distance estimate. A $\gamma$-ray study can now connect this thermal evolution to the still unconstrained non-thermal particle population and probe the acceleration efficiency and relativistic energy content of an evolved remnant.

The Fermi Large Area Telescope (Fermi-LAT), with its continuous all-sky coverage, broad energy band 
($\sim30\,\mathrm{MeV}$ to $\sim1\,\mathrm{TeV}$), and public data spanning more than a decade, is 
well suited for such an investigation \citep{atwood2009large}. We use 17 years of Pass 8 data to search for and characterise 
$\gamma$-ray emission associated with G156.2+05.7, and to examine plausible emission scenarios given its inferred middle-aged 
evolutionary state. Section~\ref{sec:data} describes the Fermi-LAT analysis, including morphological and 
spectral results. Section~\ref{sec:CO_observations} presents CO observations used to assess
possible line-of-sight contamination and CO-based morphological alternatives. Section~\ref{sec:discussion} interprets the results through multi-wavelength modelling and 
discusses the internal structure, and Section~\ref{sec:conclusion} summarises the main conclusions.

\section{Fermi-LAT data analysis}
\label{sec:data}

In this section, we characterise the $\gamma$-ray emission toward G156.2+05.7 in terms of its
spatial morphology (Section~\ref{sec:morphology}), energy spectrum
(Section~\ref{sec:energy_spectrum}), and long-term variability (Section~\ref{sec:variability}). We use the standard LAT analysis software Fermi Science Tools v2.2.0 and Fermipy 
v1.4.0 \citep{wood2017fermipy} to determine the extension and position of the extended source.
Pass 8 data collected from August 4, 2008 (MET 239557417, where MET is the mission elapsed time in seconds) to August 3,
2025 (MET 775890686) are analysed in a $15^\circ$ radius around SNR G156.2+05.7. 
Events are selected with ``Source'' class (evclass=128) and FRONT+BACK event type (evtype=3), 
applying the standard quality cuts (\texttt{DATA\_QUAL} $>0$) and (\texttt{LAT\_CONFIG} $=1$) 
and a zenith angle cut $<90^\circ$ to suppress Earth limb contamination. We work over the Fermi-LAT band (100 MeV--1 TeV), divided into 8 logarithmic bins per decade for the counts cube and exposure cube.
The Galactic interstellar diffuse background model ``gll\_iem\_v07'' and the isotropic spectral 
template ``iso\_P8R3\_SOURCE\_V3\_v1'', together with all sources from the 4FGL-DR4 catalogue 
\citep{abdollahi2022incremental,ballet2023fermi} within $15^\circ$ of the region of interest (ROI) centre, are included in the analysis 
with Fermipy. We then perform a binned likelihood fit using the MINUIT optimizer, freeing only 
the spectral parameters of catalogue sources within $10^\circ$ of the ROI centre and the normalisations 
of the two diffuse components.

\subsection{Morphological analysis}
\label{sec:morphology}

\begin{figure*}
    \centering
    \begin{subfigure}[b]{0.95\columnwidth}
        \centering
        \includegraphics[width=\columnwidth]{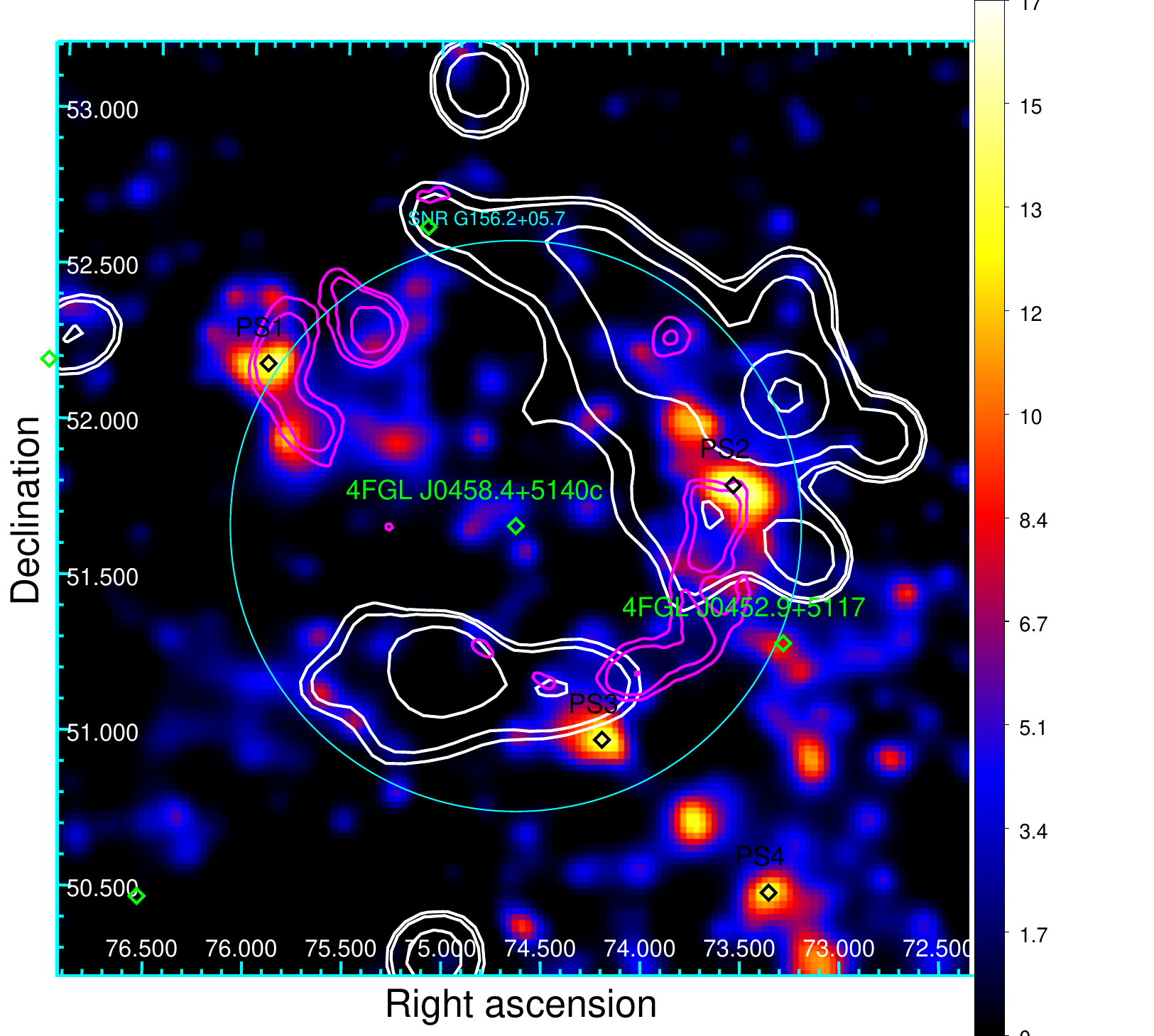}
        \caption{}
        \label{fig:ts_left}
    \end{subfigure}
    \begin{subfigure}[b]{0.95\columnwidth}
        \centering
        \includegraphics[width=\columnwidth]{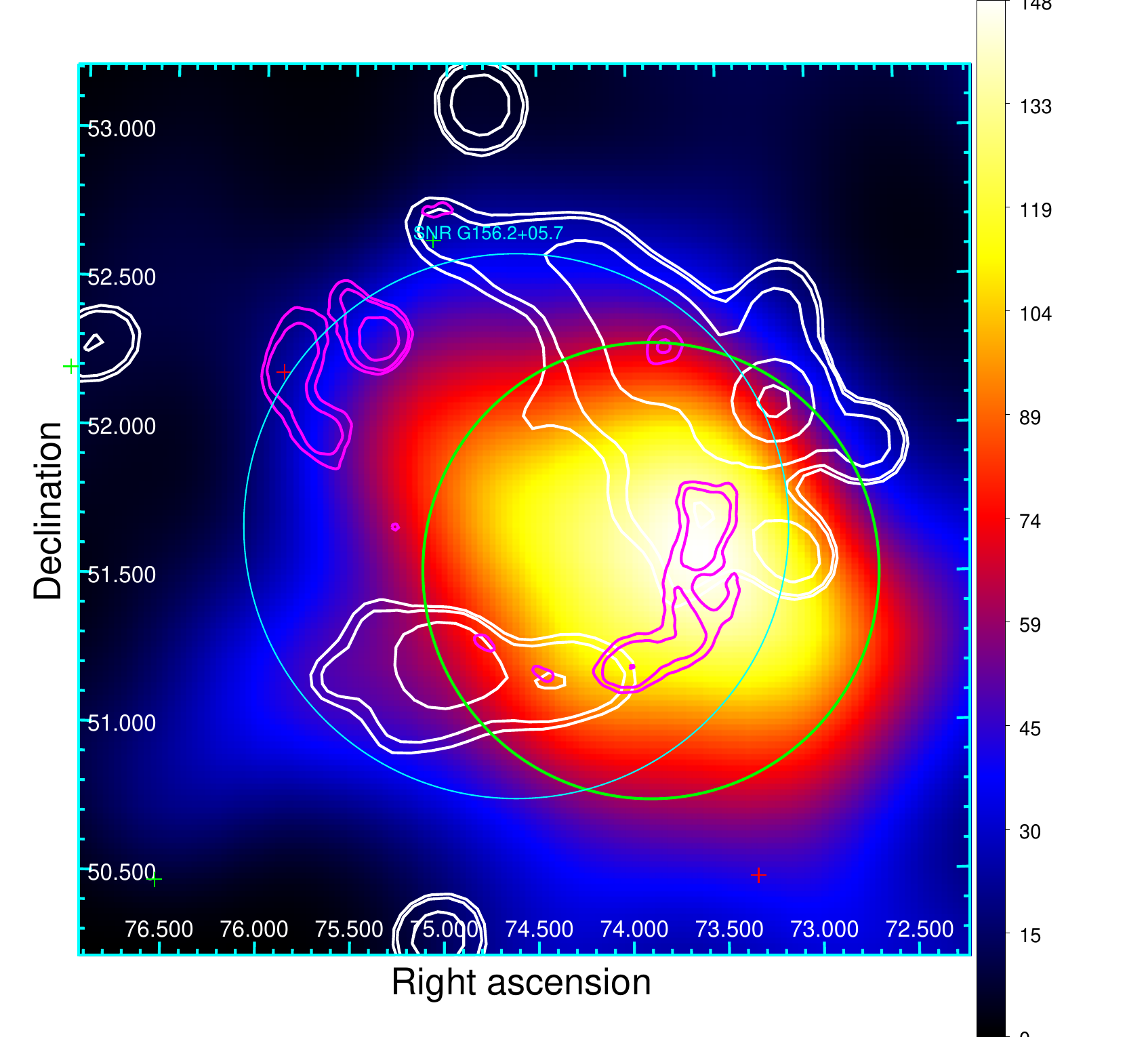}
        \caption{}
        \label{fig:ts_right}
    \end{subfigure}
    \caption{Left: $3.0^\circ \times 3.0^\circ$ TS map centred at SNR G156.2+05.7 after subtracting 4FGL-DR4 catalogue sources and diffuse
    background components. Right: $3.0^\circ \times 3.0^\circ$ TS map after subtracting 4FGL-DR4 catalogue sources
    (except 4FGL J0458.4+5140c and 4FGL J0452.9+5117) and diffuse background components.
    Green plus signs mark the 4FGL-DR4 catalogue sources shown in each panel; the two sources inside the radio shell are marked and labelled in the left panel and are not shown in the right panel; 
    white contours show the 1.4 GHz radio image from \citet{reich1992radio}; magenta contours show H$\alpha$ emission from the full-sky 
    composite of \citet{finkbeiner2003fullsky} (VTSS+SHASSA, 6-arcmin resolution); and the cyan circle traces the radio boundary. In the left panel, black plus signs labelled ps1--ps4 mark the point sources added in the multi-point-source model (Model 2). In the right panel, red plus signs mark the two supplementary point sources outside the radio shell, and the green circle marks the 68\% containment radius of the 2D Gaussian template.}
    \label{fig:ts}
\end{figure*}

In the 4FGL-DR4 catalogue, the $\gamma$-ray emission toward G156.2+05.7 is described by two sources, 4FGL J0458.4+5140c and 4FGL J0452.9+5117, both of which lie within the radio shell of the remnant. However, the test-statistic (TS) map obtained after subtracting the catalogue sources and the diffuse backgrounds still shows extended residual emission along the shell, part of which coincides with the two catalogue sources (Fig.~\ref{fig:ts}, left), indicating that the catalogue description does not fully capture the $\gamma$-ray morphology. To investigate the $\gamma$-ray emission around G156.2+05.7, we first compare the 4FGL-DR4 baseline (Model 1) with a more flexible multi-point-source model (Model 2), and then refit the region with single extended templates: a 2D Gaussian without supplementary point sources (Model 1b), a uniform disc (Model 3), and a 2D Gaussian (Model 4). We also test CO-based H$_2$ gas templates (Models 5 and 6) as an alternative morphological description. The TS map is generated in the 100 MeV--1 TeV band, including all 4FGL-DR4 sources within $15^\circ$ and the two diffuse background components.

\begin{table*}
\centering
\caption{\label{tab:model_comparison}Spatial-model comparison for the $\gamma$-ray emission toward G156.2+05.7.}
\begin{tabular}{lrrrrr}
\hline
Spatial model & $\ln\mathcal{L}$ & $k_{\rm free}$ & $\Delta\mathrm{AIC}$\textsuperscript{b} & RA, Dec (deg, J2000) & Extension\\
\hline
Model 1 (4FGL-DR4) & -2268415.7 & 10 & 0 & \makecell{$74\fdg75 \pm 0.14$, $51\fdg75 \pm 0.18$\\$73\fdg27 \pm 0.11$, $51\fdg20 \pm 0.14$} & ---\\
\hline
Model 2 (multi-point) & -2268396.2 & 26 & -7.0 & \makecell{$75\fdg91 \pm 0.07$, $52\fdg19 \pm 0.07$\\$73\fdg49 \pm 0.05$, $51\fdg80 \pm 0.03$\\$74\fdg18 \pm 0.04$, $50\fdg98 \pm 0.03$\\$73\fdg35 \pm 0.03$, $50\fdg49 \pm 0.03$} & ---\\
\hline
Model 1b (1 Gaussian) & -2268374.650 & 5 & -92.1 & $73\fdg89 \pm 0.07$, $51\fdg45 \pm 0.07$ & $0\fdg87^{+0.10}_{-0.09}$\\
\hline
Model 3 (1 disc + 2 points) & -2268361.6 & 13 & -102.2 & \makecell{$73\fdg72 \pm 0.06$, $51\fdg28 \pm 0.07$\\$75\fdg89 \pm 0.04$, $52\fdg19 \pm 0.04$\\$73\fdg34 \pm 0.04$, $50\fdg49 \pm 0.03$} & $0\fdg71^{+0.03}_{-0.04}$\\
\hline
Model 4 (1 Gaussian + 2 points) & -2268361.2 & 13 & -103 & \makecell{$73\fdg79 \pm 0.28$, $51\fdg34 \pm 0.06$\\$75\fdg89 \pm 0.04$, $52\fdg19 \pm 0.04$\\$73\fdg34 \pm 0.04$, $50\fdg49 \pm 0.03$} & $0\fdg84^{+0.09}_{-0.09}$\\
\hline
Model 5 (H$_2$~map\textsuperscript{a} + 1 point) & -2268377.2 & 6 & -85 & $75\fdg89 \pm 0.04$, $52\fdg19 \pm 0.04$ & ---\\
\hline
Model 6 (H$_2$~map\textsuperscript{a} + half-Gaussian + 2 points) & -2268368.7 & 15 & -84.2 & \makecell{$75\fdg89 \pm 0.04$, $52\fdg19 \pm 0.04$\\$73\fdg34 \pm 0.04$, $50\fdg49 \pm 0.03$} & ---\\
\hline
\end{tabular}
\vspace{4pt}

\textsuperscript{a}The H$_2$ map is derived from the $^{12}\mathrm{CO}$ $J=1\to0$ intensity map integrated over the 
velocity range 0--10~km~s$^{-1}$ in the region.

\textsuperscript{b}AIC differences are computed from unrounded likelihoods and may differ slightly from values recomputed using the rounded likelihoods listed here.

\end{table*}

Model 1 is the 4FGL-DR4 baseline retaining these two catalogue sources, both modelled as LogParabola spectra. Model 2 adds four PowerLaw point sources to that baseline, testing whether a more flexible point-source description can replace an extended component. As a cleaner test of extension, Model 1b replaces the two catalogue sources with a single 2D Gaussian. Models 3 and 4 likewise replace them with a uniform disc or a 2D Gaussian, but include two supplementary point sources outside the radio shell ($\mathrm{TS}=16.2$ and 12.2) to absorb possible nearby contamination. Finally, Model 5 combines an H$_2$ template, derived from the CO survey described in Section~\ref{sec:CO_observations}, with one supplementary point source, while Model 6 combines the H$_2$ template with a half-Gaussian template and two such point sources. The fitted positions and extensions are listed in Table~\ref{tab:model_comparison}; extensions of Models 3 and 4 are obtained with the Fermipy \texttt{extension} method, and positions of the additional point sources are obtained with \texttt{localize}. The reported $k_{\rm free}$ values count all free parameters, with five for each LogParabola catalogue source (position, normalisation, index, curvature), four for each additional PowerLaw point source (position, normalisation, index), and five for each parametric extended template (position, extension, normalisation, index). The CO-based H$_2$ template has a fixed spatial profile and contributes only its spectral parameters (normalisation and index), while the half-Gaussian in Model~6 is clipped along a fixed axis and therefore contributes the same five parameters as the full Gaussian.

For the model comparison we use the test statistic $\text{TS} = 2 (\ln \mathcal{L}_{1} - \ln \mathcal{L}_{0})$, where $\mathcal{L}_{1}$ is the maximum likelihood with the target source plus backgrounds and $\mathcal{L}_{0}$ is the background-only maximum likelihood, and the significance of an extended model relative to a point source is given by $\mathrm{TS}_{\text{ext}} = 2 (\ln \mathcal{L}_{\text{ext}} - \ln \mathcal{L}_{\text{pt}})$, with $\mathcal{L}_{\text{ext}}$ and $\mathcal{L}_{\text{pt}}$ the likelihoods under the extended and point-source hypotheses, respectively. Both extended templates improve the fit significantly over the point-source hypothesis, with $\mathrm{TS}_{\text{ext}} = 74.9$ and $75.7$ for Models 3 and 4, respectively. Using the Akaike Information Criterion \citep{akaike1974new}, defined as $\Delta\mathrm{AIC}_i = \mathrm{AIC}_i-\mathrm{AIC}_{\text{Model 1}}$ with $\mathrm{AIC}=-2\ln\mathcal{L}+2k_{\rm free}$ and negative values indicating a lower AIC than Model 1, the 2D Gaussian is slightly favoured ($\Delta\mathrm{AIC}=-103$ vs.\ $-102.2$), while Model 1b (2D Gaussian only, without extra point sources) yields $\Delta\mathrm{AIC}=-92.1$ with 5 rather than 10 free parameters, supporting the extended-source description over the two-point-source baseline. The CO-based H$_2$ templates (Models 5 and 6) improve on Model 1 but remain worse than the Gaussian. We therefore adopt the 2D Gaussian (Model 4) as the baseline spatial model.

The adopted model assumes a circular profile and an energy-independent morphology. We test both assumptions by fitting elliptical Gaussian and elliptical disc templates and by repeating the Gaussian fit using only events above 1 GeV; the numerical results are summarised in Appendix Table~\ref{tab:morphology_checks}. The best elliptical template has $R_{68}$ semi-axes of $0\fdg95$ and $0\fdg65$ at a position angle of $44\fdg75$, but the likelihood gain over the circular counterpart is only $\Delta\ln\mathcal{L}\approx3$ for two additional shape parameters and is therefore not decisive under the AIC comparison. The $>1$ GeV fit returns $R_{68}=1\fdg05^{+0.13}_{-0.12}$, larger than the full-band value, so the higher-energy data do not indicate a more compact morphology. These checks support retaining the circular 2D Gaussian for the subsequent analyses.

Although the adopted template is circularly symmetric, the residual emission is stronger in the northeast and southwest than in the northwest and southeast (Fig.~\ref{fig:ts}, right). To quantify this asymmetry, we divide the adopted 2D Gaussian template into four quadrants along the RA and Dec axes and fit all four quadrants simultaneously. The northeast and southwest quadrants have the highest detection significance (TS = 57.9 and 40.9, compared with 7.6 and 9.6 for the northwest and southeast). The two high-significance quadrants are used for the spectral comparison in Section~\ref{sec:energy_spectrum}.

\subsection{Energy spectrum}
\label{sec:energy_spectrum}
\begin{figure}
    \centering
    \includegraphics[width=\columnwidth]{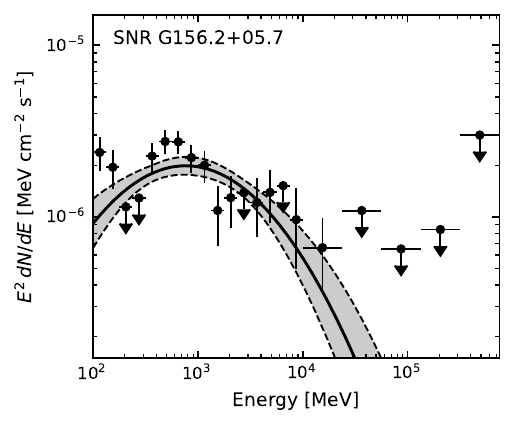}
    \caption{Spectral energy distribution (SED) of the 2D Gaussian model fitted to the $\gamma$-ray emission toward SNR G156.2+05.7 (100 MeV--1 TeV, 17 years of Fermi-LAT data). The SED points use the 21 energy bins as the joint SED fits (Section~\ref{sec:disc:sed}). Solid curves are the best-fit 
    models; dashed lines indicate the 1$\sigma$ uncertainty bands.}
    \label{fig:sco_gaussian}
\end{figure}

\begin{table*}
\centering
\caption{\label{tab:curvature_result}Spectral-curvature comparison between power-law and log-parabola models.}
\begin{tabular}{llrrr}
\hline
Type & Formula & $\ln\mathcal{L}$ & $\mathrm{TS}_\mathrm{LP}$\textsuperscript{a} & $\Delta\mathrm{AIC}$\textsuperscript{b}\\
\hline
PL & $dN/dE = \mathrm{N}_{0} (E/E_{0})^{-\Gamma}$ & -2268361.901 & -- & 0\\
LogPb & $dN/dE = \mathrm{N}_{0} (E/E_{0})^{-\Gamma-\beta\ln(E/E_{0})}$ & -2268356.187 & 11.429 & -9.429\\
\hline
\end{tabular}
\vspace{4pt}

\textsuperscript{a}$\mathrm{TS}_\mathrm{LP}$ is defined as $2(\ln\mathcal{L}_{\mathrm{LogPb}}-\ln\mathcal{L}_{\mathrm{PL}})$.

\textsuperscript{b}Differences are calculated with respect to the model with a power-law spectrum.
\end{table*}

With the best-fit 2D Gaussian spatial model fixed, we extract the spectrum assuming either a power law 
(PL; $dN/dE \propto E^{-\Gamma}$) or a log-parabola (LogPb; $dN/dE \propto E^{-\Gamma-\beta\ln(E/E_0)}$) 
over 100 MeV--1 TeV. Spectral curvature is tested using the curvature method in Fermipy, which performs 
a likelihood ratio test against a power-law hypothesis. The fitting results in Table~\ref{tab:curvature_result} 
reveal evidence for curvature, with $\mathrm{TS}_{LP} = 2(\ln\mathcal{L}_{\text{LogPb}}-\ln\mathcal{L}_{\text{PL}}) 
= 11.429$; for one additional parameter this corresponds to $\approx3.4\sigma$. The LogPb model yields 
$\Gamma = 2.101 \pm 0.089$ and $\beta = 0.187 \pm 0.081$. The integrated photon flux from 100 MeV to 
1 TeV is $1.41 \times 10^{-8}$ ph cm$^{-2}$ s$^{-1}$.

For the SED displayed in Fig.~\ref{fig:sco_gaussian}, the 100 MeV--1 TeV data are binned into the 21 logarithmic energy intervals used in the joint fits. The flux normalisations of all sources are fixed and the spectral index is set to the local slope of the global model. For bins with TS $< 4.0$ ($<2\sigma$), the 95\% confidence upper limit is reported. The resulting SED is shown together with the global best-fit spectrum. The curved spectrum may reflect the superposition of multiple emission components, which we explore through multi-wavelength modelling in Section~\ref{sec:disc:sed}.

Section~\ref{sec:morphology} showed that the $\gamma$-ray signal is concentrated in the northeast and southwest quadrants ($\mathrm{TS}=57.9$ and $40.9$), which we now compare spectrally. Fitting the same 2D-Gaussian template to each of these two quadrants separately, curvature is preferred in both: $\mathrm{TS}_{\mathrm{LP}} = 23.2$ and $9.4$ ($\approx 4.8\sigma$ and $\approx 3.1\sigma$, one additional parameter). The northeast curvature parameter reaches its upper bound ($\beta=1$) and is therefore only a lower limit; Table~\ref{tab:comparison} summarises the best-fit parameters and curvature statistics.

\begin{table*}
\centering
\caption{\label{tab:comparison}Spectral fits for the two high-significance quadrants of the 2D Gaussian template. Both are described by log-parabola models.}
\begin{tabular}{llcccc}
\hline
Quadrant & Model & $\Gamma$ & $\beta$ & $\mathrm{TS}_\mathrm{LP}$ & TS\\
\hline
northeast & LogPb & $2.715\pm0.268$ & $\ge 1.00$\textsuperscript{a} & 23.21 & 57.89\\
southwest & LogPb & $2.177\pm0.211$ & $0.121\pm0.115$ & 9.38 & 40.90\\
\hline
\end{tabular}
\vspace{4pt}

\textsuperscript{a}The northeast $\beta$ is at its upper bound; the tabulated value is therefore a lower limit.

\end{table*}

\begin{figure}
    \centering
    \includegraphics[width=\columnwidth]{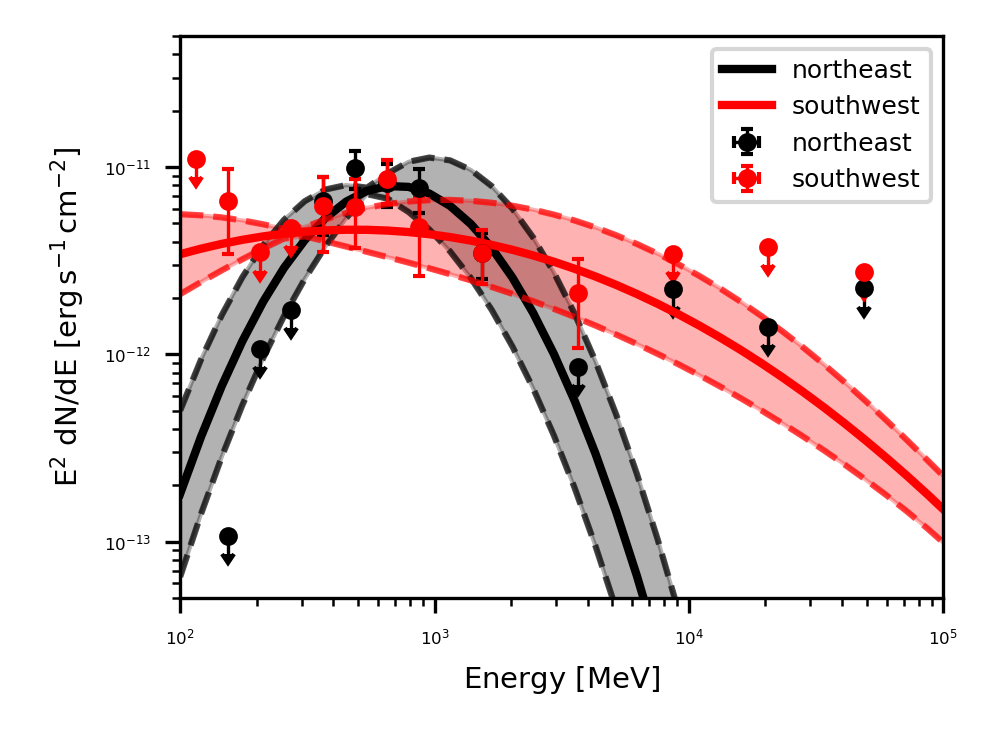}
    \caption{SEDs of the two high-significance quadrants, northeast (black) and southwest (red). Solid curves are the best-fit
    models; dashed lines indicate 1$\sigma$ uncertainty bands.}
    \label{fig:plot_two_seds}
\end{figure}

Fig.~\ref{fig:plot_two_seds} shows the SEDs of the northeast and southwest quadrants extracted from the same template. Both are described by a log-parabola, but the northeast quadrant is softer, with $\Gamma=2.71\pm0.27$ at 1 GeV, than the southwest quadrant, which has $\Gamma=2.18\pm0.21$ and only mild curvature ($\beta=0.12\pm0.12$). The spectral indices differ by $\Delta\Gamma=0.54\pm0.34$, i.e.\ at the $\approx1.6\sigma$ level and hence not significantly; the apparent difference in curvature is driven by the lower bound on $\beta$ in the northeast quadrant, so the two SEDs are consistent with a common spectral shape within the present statistics. We defer the physical interpretation of this quadrant-level comparison to Section~\ref{sec:disc:clues}. The global spectrum provides the input for the multi-wavelength modelling in Section~\ref{sec:disc:sed}, and, because all spectra above are time-averaged over the full 17-year span, the assumption of steady emission underlying this analysis is tested in Section~\ref{sec:variability}.

\subsection{Flux variability}
\label{sec:variability}

\begin{figure}
    \centering
    \includegraphics[width=\columnwidth]{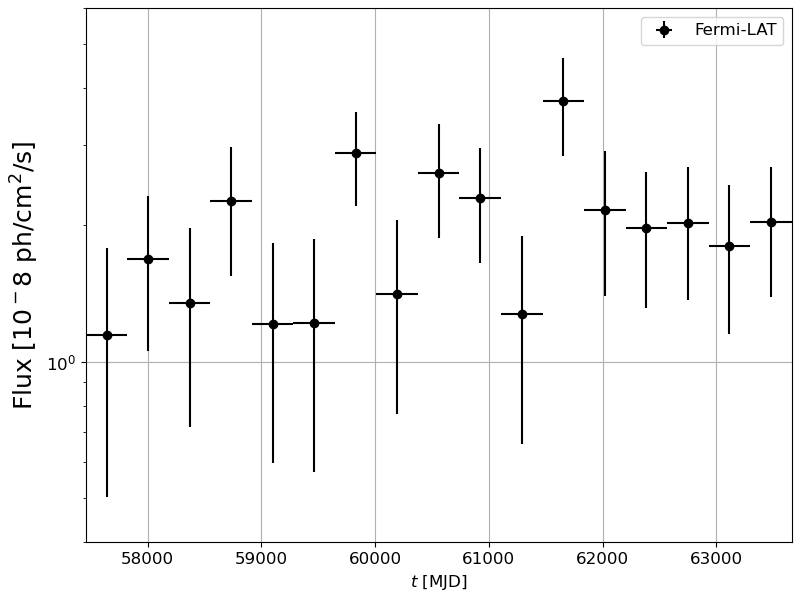}
    \caption{Light curve of the $\gamma$-ray emission toward SNR G156.2+05.7 in the 100 MeV--1 TeV band with 1-year time bins. Black points show flux measurements with error bars.}
    \label{fig:lc}
\end{figure}

Given the extended spatial model established in Section~\ref{sec:morphology}, the source is expected to show no long-term variability. We test this expectation by constructing a 1-year binned light curve with the adopted 2D Gaussian model (Fig.~\ref{fig:lc}) in the 100 MeV--1 TeV band with the lightcurve method in Fermipy; the 17 time bins each cover approximately one year. Following \citet{nolan2012fermi}, a variability index 
$\mathrm{TS}_{\text{var}} > 34.0$ is required to claim 99\% confidence temporal variability for 17 bins. 
The measured $\mathrm{TS}_{\text{var}} = 16.1$ indicates no significant variability, consistent with 
steady emission from an extended source.

\section{CO observations}
\label{sec:CO_observations}

\begin{figure}
    \centering
    \includegraphics[width=\columnwidth]{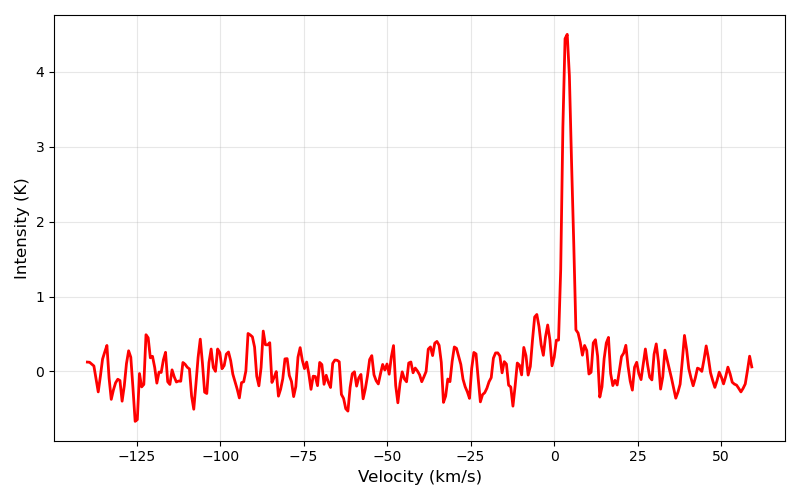}
    \caption{Velocity channel map of the region.}
    \label{fig:velocity_channel_map}
\end{figure}
\begin{figure}
    \centering
    \includegraphics[width=\columnwidth]{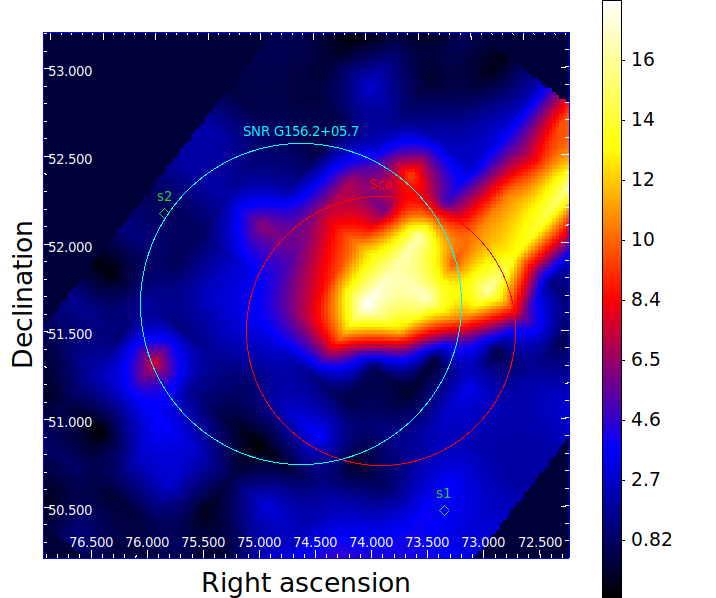}
    \caption{$^{12}\text{CO}$ $J=1\to0$ intensity map integrated over 0--10 km s$^{-1}$ in the region.}
    \label{fig:co_map}
\end{figure}

We use $^{12}\text{CO}$ $J=1\to0$ data from the whole-Galaxy CO survey 
\citep{dame2001milky}; we specifically use the survey subregion labelled ``DHT18'' (hereafter the DHT18 tile) 
to characterise the molecular gas toward G156.2+05.7. 
The DHT18 survey covers the northern Galactic plane with $0.25^{\circ}$ spatial sampling and a velocity 
resolution of $1.3\,\text{km}\,\text{s}^{-1}$ after interpolation. The angular resolution (full width at half-maximum, FWHM) is 
$\sim 8.8^{\prime}$, and the typical rms noise is $\sim 0.15\,\text{K}$ per channel.

We extract a subcube spanning $l = 154.5^{\circ}$--$157.5^{\circ}$ and $b = 3.5^{\circ}$--$6.5^{\circ}$ 
using the spectral-cube package (\url{https://spectral-cube.readthedocs.io}). 
Fig.~\ref{fig:velocity_channel_map} shows the velocity channel map. The molecular emission is concentrated 
at low velocities with respect to the local standard of rest (LSR); we adopt the $0$--$10\,\text{km}\,\text{s}^{-1}$ slab for further analysis. At these 
velocities, the gas is likely associated with the nearby Cepheus flare cloud complex \citep{kun2008star} or the local Orion arm. Applying 
the Galactic rotation curve of \citet{brand1993velocity}, adopting the recent
Galactic parameters $R_0 = 8.34\,\text{kpc}$ and 
$\Theta_0 = 240\,\text{km}\,\text{s}^{-1}$ \citep{reid2014trigonometric}, gives a kinematic distance of roughly $0.5$--$1.0\,\text{kpc}$, 
placing the cloud in the foreground of G156.2+05.7.

Integrated intensity maps (moment 0) constructed over $0$--$10\,\text{km}\,\text{s}^{-1}$ 
(Fig.~\ref{fig:co_map}) reveal a prominent molecular complex across the field.

To estimate the mean molecular gas density, we convert the velocity-integrated CO intensity,
$W_{\rm CO}=\int T_{\rm mb}(v)\,dv$, into the H$_2$ column density via the standard
$X_{\rm CO}$ factor, $N(\mathrm{H}_2) = X_{\rm CO}\,W_{\rm CO}$, adopting
$X_{\rm CO}=2\times 10^{20}\,\mathrm{cm^{-2}\,(K\,km\,s^{-1})^{-1}}$\citep{bolatto2013cotoh2}.
Converting the mean integrated intensity $\langle W_{\rm CO}\rangle = 8.87\,\mathrm{K\,km\,s^{-1}}$
with a line-of-sight depth $L\simeq 2R$ at $D\simeq 0.5$--$1.0\,\mathrm{kpc}$ yields
$n(\mathrm{H}_2) \sim 27\text{--}55\,\mathrm{cm^{-3}}$.

The molecular morphology partially overlaps the $\gamma$-ray emission discussed in Section~\ref{sec:data}. 
Because the 0--10~km~s$^{-1}$ gas lies in the foreground, we treat it as a possible line-of-sight contaminant and as a morphological alternative in the spatial-model tests, rather than as the physical target of the emission.

\section{Discussion}
\label{sec:discussion}

The Fermi-LAT analysis reveals extended $\gamma$-ray emission toward G156.2+05.7 with evidence for spectral 
curvature ($\mathrm{TS}_{LP} = 11.429$, $\approx 3.4\sigma$) and no temporal variability 
($\mathrm{TS}_{\text{var}} = 16.1$). Below we interpret these results, moving from the emission site
to the emission mechanism, and then to the target-density and internal-structure implications.

\subsection{Spatial morphology and the emission site}
\label{sec:disc:site}

The extent and centroid of the $\gamma$-ray source provide the most direct constraint on where the emission
originates. The 2D Gaussian template gives $R_{68} \sim 0.84^{\circ}$. Adopting the 
reference distance $d = 2.7$ kpc (Section~\ref{sec:intro}), this translates to a physical radius of $\sim 40$ pc
(diameter $\sim 79$ pc), compatible with the
$\sim 0.9^\circ$ radio shell within uncertainties. The $\gamma$-ray centroid is offset from the radio centre by
$\sim 0.6^\circ$. Because the radio shell is faint and asymmetric, and the 2D Gaussian is fitted to
non-uniform $\gamma$-ray emission (Section~\ref{sec:morphology}), we do not interpret this offset as direct
evidence for a physical displacement between the two emission regions. The origin and spectral
signatures of this asymmetry are examined in Section~\ref{sec:disc:clues}.

The morphological tests (Table~\ref{tab:model_comparison}) show that the $0$--$10$ km s$^{-1}$ foreground CO template does not reproduce the $\gamma$-ray morphology, whereas a 2D Gaussian provides a better description. These clouds (Section~\ref{sec:CO_observations}) lie in the foreground and partially overlap the $\gamma$-ray region; we therefore treat them as a line-of-sight contaminant rather than as the physical target of the emission.

\subsection{Joint leptonic--hadronic modelling}
\label{sec:disc:sed}

To model the broadband emission we fit the radio data and the Fermi-LAT spectrum jointly. We use the global spectrum because it has higher statistics than the individual quadrant spectra of Section~\ref{sec:energy_spectrum}, which serve only to characterise the internal structure of the remnant (Section~\ref{sec:disc:clues}). The energy-spectrum
analysis (Section~\ref{sec:energy_spectrum}) finds evidence for curvature, with the
log-parabola model preferred over a single power law at the $\approx 3.4\sigma$ level. This
motivates a direct three-way comparison of emission hypotheses:
\begin{enumerate}
    \item a single-leptonic model (\texttt{lep}), in which one electron
    population produces both radio synchrotron and IC $\gamma$-ray emission;
    \item a leptonic--hadronic model (\texttt{lep\_had}), in which
    electrons account for the radio and part of the $\gamma$-ray emission while proton--proton
    interactions produce the remaining $\pi^0$-decay $\gamma$-ray emission; and
    \item a hadronic-plus-synchrotron model (\texttt{hadsyn}), in which
    protons produce all of the $\gamma$-ray emission and a separate electron population
    supplies the radio synchrotron.
\end{enumerate}

We perform joint modelling of the Fermi-LAT spectrum and radio synchrotron
constraints \citep{xu2007polarization} using the \textsc{Naima} package
\citep{zabalza2015naima} with Markov Chain Monte Carlo (MCMC) sampling via \textsc{emcee} \citep{foreman-mackey2013emcee}. We run $128$ walkers for $500$ burn-in steps followed by $5000$ production steps per walker, and quote posterior medians with 16--84\% credible intervals. The \texttt{lep} and \texttt{lep\_had} models include IC scattering on three seed photon fields: the cosmic microwave background (CMB)
($T = 2.73$ K, $u = 0.26$ eV cm$^{-3}$), far-infrared dust emission
($T = 30$ K, $u = 0.1$ eV cm$^{-3}$), and near-infrared starlight
($T = 3000$ K, $u = 0.1$ eV cm$^{-3}$) \citep{porter2006inverse}. In \texttt{hadsyn}
the electron population is treated as synchrotron-only, so that \texttt{hadsyn} is the
no-IC counterpart of \texttt{lep\_had} and the model comparison isolates the leptonic
GeV channel. For the fiducial
fits we adopt $B = 14\,\mu\mathrm{G}$, the minimum-energy estimate used by
\citet{pannuti2004asca} and based on \citet{reich1992radio}; the sensitivity of the
adopted model to this assumption is tested after the model comparison below.
\citet{uchida2012suzaku} showed that the X-ray emission is dominated by thermal
components from shock-heated ISM, with negligible non-thermal synchrotron, so we do
not include X-ray data in the joint fits.

Injected particle spectra follow a power law with an exponential cutoff,
\begin{equation}
\frac{dN}{dE} = N_0 \left(\frac{E}{E_0}\right)^{-\alpha}
\exp\left(-\frac{E}{E_{\mathrm{cut}}}\right),
\end{equation}
where $N_0$ is the particle-number normalisation in GeV$^{-1}$ (the number of particles per unit energy, integrated over the emitting volume), $E_0$ the reference energy (fixed at $1$ TeV in the fits), $\alpha$ the spectral
index, and $E_{\mathrm{cut}}$ the cutoff energy. The electron component has free
$N_{0,e}$, $\alpha_e$, and $E_{e,\mathrm{cut}}$. In the two models containing protons,
the hadronic component has free $A_{pp}$ and $\alpha_p$, while $E_{p,\mathrm{cut}}$ is
fixed at $1$ TeV because the Fermi-LAT data do not constrain it
(Fig.~\ref{fig:had_cutoff_compare}). Here $A_{pp} \equiv N_{0,p}\,n$ is the
product constrained by the SED, in units of cm$^{-3}$ GeV$^{-1}$
(Table~\ref{tab:model_parameters}); this choice avoids treating $n$ and $N_{0,p}$ as
independently constrained quantities, and the physical density is introduced only in
the density-dependent interpretation below.

All models are fitted to the identical 25-point Fermi-LAT+radio dataset at the
reference distance. For the Fermi-LAT part, we use the native grid of 8 logarithmic bins
per decade employed for the counts and exposure cubes. The 16 bins below 10 GeV are retained,
and the bins above 10 GeV are merged into five bins, giving 21 energy bins in total.
Fig.~\ref{fig:sco_gaussian} shows the same Fermi-LAT data points used in the
fits; the remaining points are the radio measurements. Normalisations, amplitudes, and
cutoff energies are sampled uniformly in
$\log_{10}$, whereas spectral indices are sampled uniformly in their native
parameters. The priors are broad and uninformative, and no additional physical
priors are imposed.

The fitted electron cutoff
($E_{e,\mathrm{cut}} \approx 1.2$ TeV) keeps the maximum synchrotron photon energy
below the X-ray band, consistent with the thermal-dominated X-ray spectrum, while the
steep proton population ($\alpha_p \approx 3.2$) with $E_{p,\mathrm{cut}} = 1$ TeV
fixed is consistent with the LHAASO non-detection \citep{cao2024first}; no dedicated
upper limit for this source is provided in that catalogue, so the non-detection is
only qualitative.

\begin{figure*}
    \centering
    \begin{subfigure}[b]{0.48\textwidth}
        \includegraphics[width=\columnwidth]{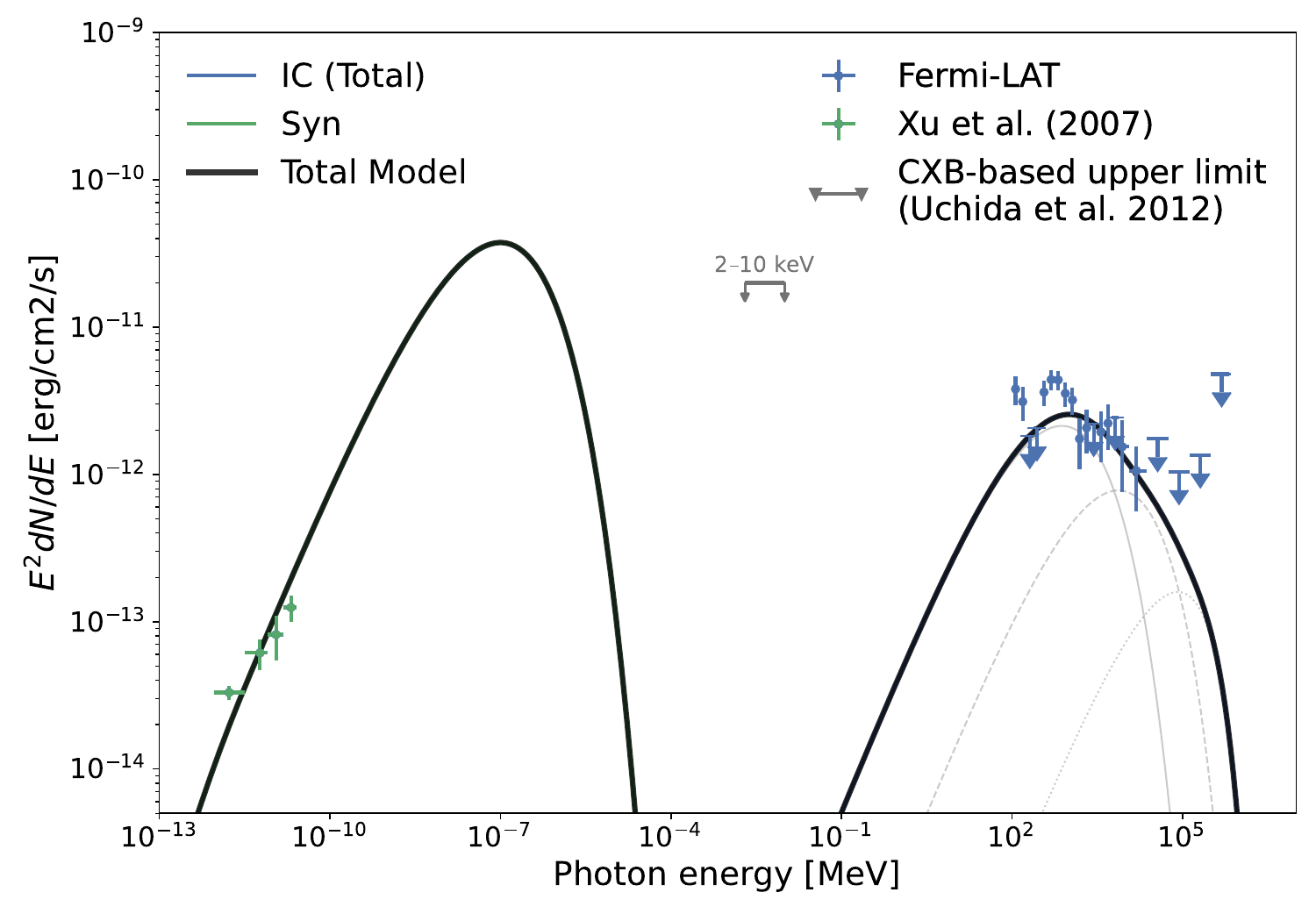}
        \caption{}
        \label{fig:lep_combined_fit_and_plot}
    \end{subfigure}
    \begin{subfigure}[b]{0.48\textwidth}
        \includegraphics[width=\columnwidth]{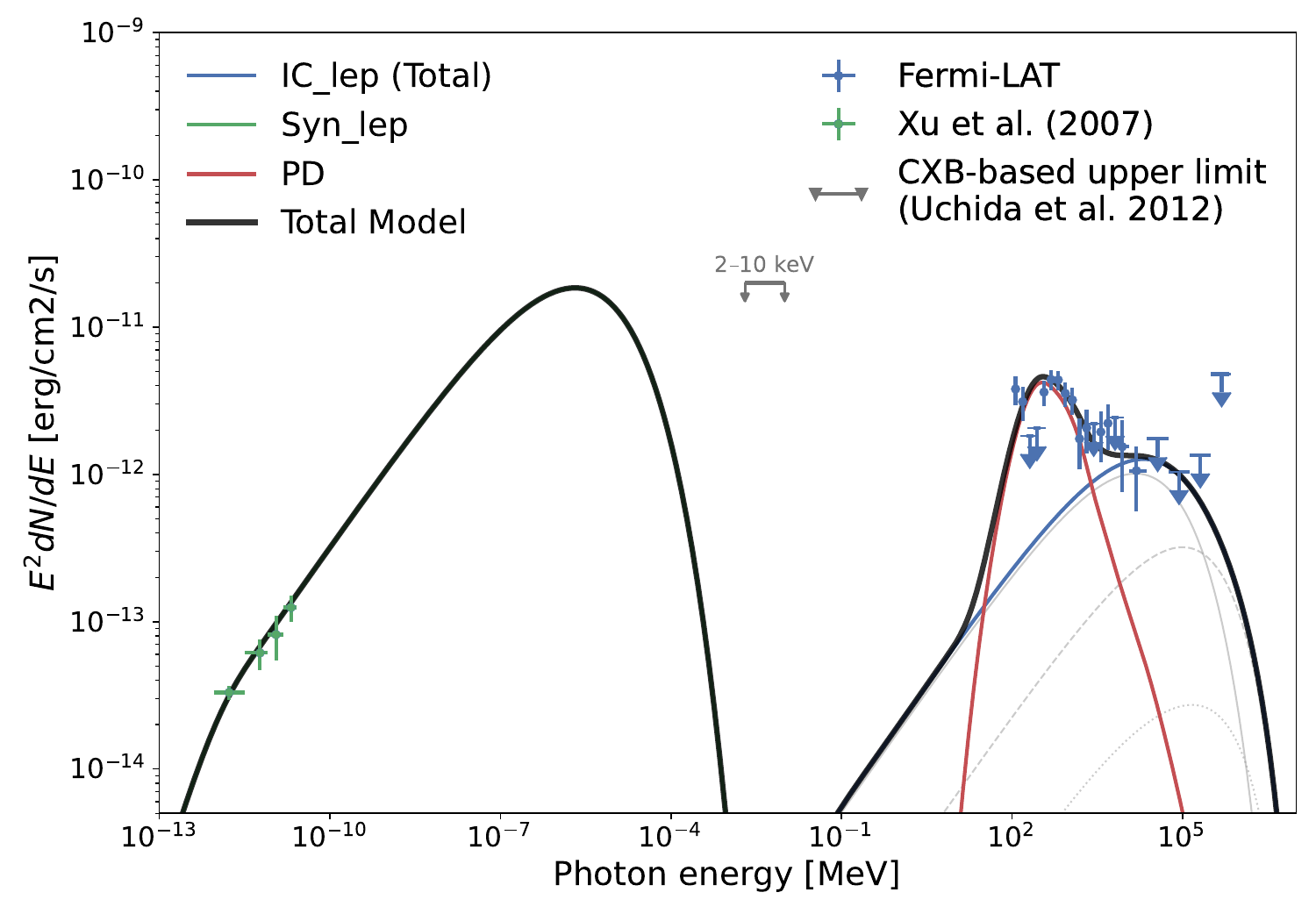}
        \caption{}
        \label{fig:lephad_combined_fit_and_plot}
    \end{subfigure}
\newline
    \begin{subfigure}[b]{0.48\textwidth}
        \centering
        \includegraphics[width=\columnwidth]{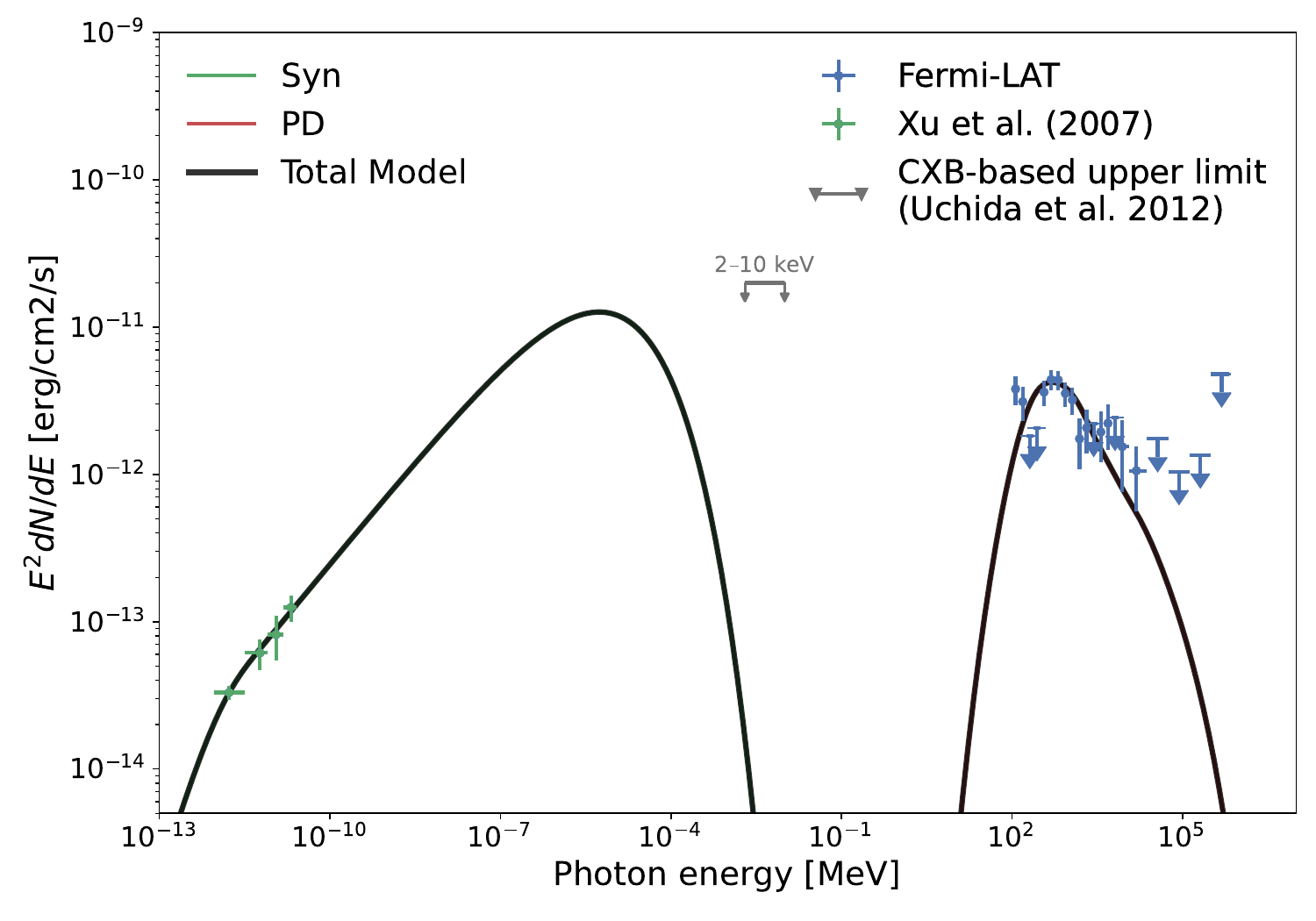}
        \caption{}
        \label{fig:hadsyn_combined_fit_and_plot}
    \end{subfigure}
    \caption{Joint fits to the radio data \citep{xu2007polarization} and the Fermi-LAT spectrum are shown. 
    The grey horizontal line with downward arrows at both ends spans $2$--$10$ keV and marks the X-ray upper limit based on the cosmic X-ray background (CXB) \citep{uchida2012suzaku}, which is 
    overplotted for comparison and is not included in the fit. Panels show the single-leptonic (lep), 
    leptonic--hadronic (lep\_had), and hadronic-plus-synchrotron (hadsyn) models, all fitted to the identical 
    Fermi-LAT+radio dataset. The \texttt{lep} and \texttt{lep\_had} models include IC scattering on three seed
    photon fields (cosmic microwave background, CMB; far-infrared, FIR; and near-infrared, NIR) together with synchrotron emission at $B = 14\,\mu\mathrm{G}$
    \citep{pannuti2004asca}; in the \texttt{hadsyn} model the primary-electron population is treated as
    synchrotron-only, so that its GeV emission is purely hadronic and \texttt{hadsyn} is the no-IC counterpart of
    \texttt{lep\_had}. Thick solid curves show the best-fit total model spectra and the labelled emission
    components; thin grey curves show the individual IC components of the \texttt{lep} and \texttt{lep\_had} models.}
    \label{fig:combined_fit_and_plot}
\end{figure*}

For model comparison, we compute the Bayesian information criterion at
the maximum-likelihood parameters,
$\mathrm{BIC} = -2\ln\mathcal{L}_{\max} + k\ln N$, where $N=25$ is the
number of data points and $k=3$ (\texttt{lep}) or $5$
(\texttt{lep\_had}, \texttt{hadsyn}) free parameters; fixed parameters are
not counted in $k$. We quote $\Delta\mathrm{BIC}$ relative to the best-fitting
\texttt{lep\_had} model, and lower BIC is preferred. The resulting comparison
is summarised in Fig.~\ref{fig:combined_fit_and_plot}
and Table~\ref{tab:model_parameters}. The pure-leptonic model is strongly disfavoured
($\Delta\mathrm{BIC} = 30.2$ relative to the leptonic--hadronic model), because a single
electron population cannot simultaneously reproduce the radio slope and the curved Fermi-LAT
spectrum; the \texttt{lep} model is a deliberately minimal leptonic description, and this
comparison does not exclude leptonic models with a second electron component. The hadronic-plus-synchrotron model improves substantially over the pure
leptonic case (BIC $=45.40$ versus $68.84$), but remains disfavoured relative to the
leptonic--hadronic model ($\Delta\mathrm{BIC}=6.8$); its electron component is poorly
constrained because it contributes only to the radio band. Because \texttt{lep\_had}
and \texttt{hadsyn} have the same number of free parameters, this $\Delta\mathrm{BIC}$
corresponds to $2\Delta\ln\mathcal{L}\approx6.8$, a maximum-likelihood difference of
only $\sim3.4$, so the preference should not be regarded as decisive. The
leptonic--hadronic model gives the lowest BIC and is adopted as the working
baseline for the density-dependent interpretation below, but the
hadronic-plus-synchrotron model is carried forward as a systematic alternative.
The ranking between the two models is also conditional on the assumed proton
cutoff ($E_{p,\mathrm{cut}}=1$ TeV, Section~\ref{sec:disc:sed}). The robust
conclusion is that the data favour a hadronic contribution over the pure-leptonic
model; whether the low-energy GeV emission is dominated by inverse Compton or
$\pi^0$-decay, and the associated proton energy budget, remain model-dependent
and depend on the adopted assumptions ($E_{p,\mathrm{cut}}=1$ TeV,
$B=14\,\mu\mathrm{G}$, a one-zone emitting region, and $d=2.7$ kpc).

\begin{table}
\centering
\caption{\label{tab:model_parameters}Posterior medians and model comparison for the three Fermi-LAT+radio SED models.}
\begin{tabular}{lccc}
\hline
 & lep & lep\_had & hadsyn\\
\hline
BIC & 68.84 & 38.62 & 45.40\\
$\Delta\mathrm{BIC}$ & 30.2 & 0 & 6.8\\
$\alpha_e$ & 1.138 & 1.930 & 2.060\\
$E_{e,\mathrm{cut}}$ (TeV) & 0.215 & 1.18 & 5.6\textsuperscript{a}\\
$\alpha_p$ & -- & 3.23 & 2.83\\
$E_{p,\mathrm{cut}}$ (TeV) & -- & 1.0 (fixed) & 1.0 (fixed)\\
$A_{pp}$ (cm$^{-3}$ GeV$^{-1}$) & -- & $2.88\times10^{43}$ & $2.72\times10^{44}$\\
\hline
\end{tabular}
\vspace{4pt}

The columns correspond to the single-leptonic (\texttt{lep}),
leptonic--hadronic (\texttt{lep\_had}), and hadronic-plus-synchrotron
(\texttt{hadsyn}) models. Parameter values are posterior medians; the
16--84\% credible intervals and maximum-likelihood values are provided in
the online supplementary material. The $\Delta\mathrm{BIC}$ entries compare
\texttt{lep} and \texttt{hadsyn} with \texttt{lep\_had}.

\textsuperscript{a}The \texttt{hadsyn} electron cutoff is poorly
constrained; its 16--84\% credible interval is
$\approx0.36$--$84.8$ TeV.

\end{table}

Because the magnetic-field strength is a key but poorly constrained parameter, we next test the sensitivity of this choice to the assumed $B = 14\,\mu\mathrm{G}$. We repeat the adopted leptonic--hadronic fit for $B = 5$, $14$, and $30\,\mu\mathrm{G}$, spanning the range from a weak ambient field to a moderately amplified one; details are given in Appendix~\ref{sec:app:bfield}. The BIC, evaluated at the maximum-likelihood point of each fit, decreases toward weaker fields ($\Delta\mathrm{BIC}\approx3$ between $B=5$ and $14\,\mu\mathrm{G}$, and $\approx9$ between the extremes), but the $B=5$ and $B=14$ fits are not significantly separated; we therefore retain the reference minimum-energy value $B=14\,\mu\mathrm{G}$ rather than selecting the numerically preferred field, since the scan is intended to test robustness and not to determine $B$. The hadronic component persists at all three field strengths. Its influence on the density-dependent interpretation is discussed in Section~\ref{sec:disc:scenarios} and Appendix~\ref{sec:app:bfield}. Corner plots and posterior result files for the models discussed in the text are provided in the online supplementary material.

\subsection{Effective target density and the energy budget}

\label{sec:disc:scenarios}

In the adopted leptonic--hadronic model
(Section~\ref{sec:disc:sed}), the hadronic normalisation enters only through the
product $A_{pp}=N_{0,p}\,n$, where $n$ is the uniform effective target density
(of hydrogen nuclei) of the one-zone emission region. Because the SED constrains
only this product, $N_{0,p}$ and $n$ remain degenerate. With $A_{pp}$ fixed at its fitted
value, the proton energy required to reproduce the observed $\gamma$-ray emission then
varies with the assumed effective density $n$.

As the most tenuous representative case, we adopt the pre-shock ambient
value $n = 0.036$ cm$^{-3}$, obtained by scaling the Suzaku-derived ambient
density of \citet{uchida2012suzaku},
$n_0 = 0.088\,(d/1.1\,\mathrm{kpc})^{-1}$ cm$^{-3}$ (a hydrogen density), to the
reference distance $d = 2.7$ kpc. We use this value to test whether a single
emission region at the density of the remnant interior can supply the required
protons.

\begin{figure}
    \centering
    \includegraphics[width=\columnwidth]{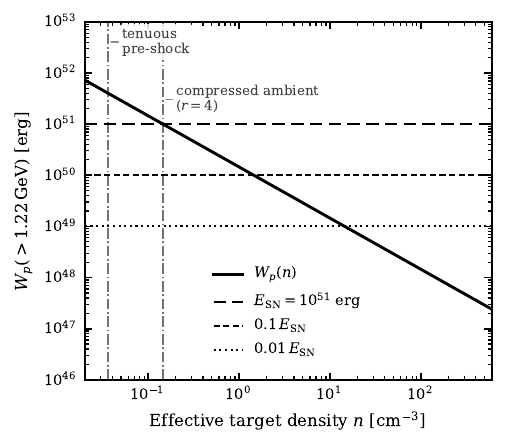}
    \caption{Proton energy above $1.22$ GeV as a function of the assumed
    effective target (hydrogen) density $n$ of the one-zone fit, at the reference
    distance with $A_{pp}$ fixed at the value adopted in
    Section~\ref{sec:disc:sed}. The dashed, dotted, and finely dotted lines mark
    the canonical $10^{51}$ erg explosion energy and $10$ and $1$ per cent of it.
    The vertical
    dash-dotted lines mark the tenuous pre-shock value
    ($n = 0.036$ cm$^{-3}$), the same gas at the maximal strong-shock compression
    ($n = 0.14$ cm$^{-3}$; Section~\ref{sec:disc:scenarios}).}
    \label{fig:wp_vs_n}
\end{figure}

The electron and proton energies are obtained with the \textsc{Naima}
functions \texttt{compute\_We} and \texttt{compute\_Wp}, which integrate
$W=\int E\,(dN/dE)\,dE$ over the posterior-median particle distributions of the
adopted fit, above the quoted threshold energies ($1$ GeV for electrons and
$1.22$ GeV for protons). Because the SED constrains the product
$A_{pp}=N_{0,p}\,n$, all density-dependent quantities follow by simple scaling
at fixed $A_{pp}$; for a fixed observed SED, $A_{pp}\propto d^2$ and hence
\begin{equation}
W_p(>1.22\,\mathrm{GeV}) \simeq
1.4\times10^{50}\left(\frac{n}{1\,\mathrm{cm^{-3}}}\right)^{-1}
\left(\frac{d}{2.7\,\mathrm{kpc}}\right)^{2}\mathrm{erg}.
\label{eq:wp_n}
\end{equation}
It is convenient to express this as a fraction of the canonical explosion
energy, $\varepsilon_{\rm cr}\equiv W_p/E_{\rm SN} \simeq
0.14\,(n/1\,\mathrm{cm^{-3}})^{-1}(d/2.7\,\mathrm{kpc})^{2}$. With $n$ left
free, we adopt a representative efficiency of about $10$ per cent
\citep{dermer2013gamma} as a benchmark for viable densities, although the
efficiency may be lower in this middle-aged remnant
\citep{zhang2019electron}. At the tenuous pre-shock density
($n = 0.036$ cm$^{-3}$), equation~(\ref{eq:wp_n}) gives
$W_p(>1.22\,\mathrm{GeV}) \sim 4\times10^{51}$ erg, about four times the
canonical $\sim 10^{51}$ erg explosion energy; for the adopted leptonic--hadronic
decomposition and $E_{p,\mathrm{cut}} = 1$ TeV, a single emission region at
this density therefore cannot supply the required total proton energy even
for an acceleration efficiency of unity. Assuming the standard compression
factor of four for a strong non-radiative shock, the post-shock density is
$n = 0.14$ cm$^{-3}$. Taking this as an optimistic effective target density
of the one-zone emission region gives $W_p \simeq E_{\rm SN}$,
corresponding to an order-unity acceleration efficiency and therefore well
above the $10$ per cent benchmark. Fig.~\ref{fig:wp_vs_n} summarises the
resulting density dependence of the energy requirement.

Within the adopted $1.7$--$3.0$ kpc range, the lower limit
$d = 1.7$ kpc is the most favorable case. At fixed $n$, $W_p \propto d^{2}$
and therefore $W_p \simeq 1.6\times10^{51}$ erg, still above $E_{\rm SN}$.
If the tenuous density is instead scaled as $n\propto d^{-1}$, following the
scaling implied by the Suzaku/Sedov fit of \citet{uchida2012suzaku}, then
$W_p \propto d^{3}$ and $W_p \simeq 1.0\times10^{51}$ erg at $d = 1.7$ kpc,
of order $E_{\rm SN}$. The nearest distance is thus the limiting case,
requiring an efficiency approaching unity within the adopted scalings.

The tenuous single-zone case is therefore energetically disfavoured, and
the required effective target density must exceed that of the remnant interior.
These energy requirements are conditional on the adopted leptonic--hadronic
decomposition, $E_{p,\mathrm{cut}}=1$ TeV, $B=14\,\mu\mathrm{G}$, and the one-zone
approximation. If the proton-dominated alternative is regarded as viable, its larger
$A_{pp}$ (Table~\ref{tab:model_parameters}) would make the energy budget more severe;
this does not, however, establish the leptonic--hadronic model as uniquely correct. We
therefore regard additional dense gas or re-acceleration as a plausible resolution of
the energy budget rather than a unique inference from the SED. The implications of
this density requirement are examined in Section~\ref{sec:disc:clues}.

\subsection{Morphological clues: asymmetry and optical filaments}
\label{sec:disc:clues}

The density requirement derived in
Section~\ref{sec:disc:scenarios} motivates an interpretation of the spatially
asymmetric $\gamma$-ray emission. As established in Sections~\ref{sec:morphology}
and~\ref{sec:energy_spectrum}, the northeast and southwest quadrants are the two
bright quadrants and both show evidence for curvature, whereas the remaining
quadrants are only marginally detected. We
therefore examine whether this established asymmetry can reflect localised
density or shock-environment variations, such that a significant part of the
hadronic emission arises where the shock interacts with denser gas. The remnant
may have swept up and rarefied much of its surroundings while continuing to
interact with a denser, inhomogeneous medium, so that the emitting volume mixes
a tenuous interior with denser gas. Relative to the $\gamma$-ray centroid, the enhanced
directions correspond in azimuth to the H$\alpha$ filaments traced by the
magenta contours in Fig.~\ref{fig:ts}, rather than strictly coinciding with them
spatially; given the offset between the $\gamma$-ray centroid and the radio shell centre
(Section~\ref{sec:disc:site}), this directional correspondence is the relevant
comparison. This is also where a positional correspondence with discrete clumps
would be expected. If the radiative filaments trace dense gas in discrete
clumps, rather than a single molecular cloud, those clumps can provide the
localised target density required by the energy-budget argument without
requiring the entire emitting volume to be dense. The possible association is
therefore between the directions of the two high-significance quadrants
and the locations where the shock encounters such clumps.
Re-acceleration of pre-existing Galactic cosmic rays in crushed clouds can
further reduce the demand for freshly accelerated protons
\citep{uchiyama2010gammaray,cardillo2016supernova}, as also invoked for the
GeV-bright, TeV-quiet remnants W44, IC~443, and W41
\citep{cardillo2016supernova,murase2026gas}. The present data cannot establish
this association unambiguously, nor decide whether fresh acceleration or
re-acceleration dominates; both require tighter target-density constraints.

We interpret this asymmetric enhancement, together with the marginal spectral difference reported in Section~\ref{sec:energy_spectrum},
as tentative evidence for localised density or shock-environment
inhomogeneities. Such interactions may enhance target densities or modify local
acceleration conditions, affecting both hadronic and leptonic emission, for
example through turbulence-amplified magnetic fields \citep{bell2004turbulent}.
On larger scales, expansion into an ambient density gradient could additionally
enhance the radio synchrotron emission on one side by compressing the magnetic
field more efficiently \citep{orlando2007origin,reynolds2008supernova}, while
the $\gamma$-ray emission traces the broader post-shock particle distribution
\citep{reynolds2008supernova}.
Given the limited photon statistics
in the quadrant fits, these spectral differences should be interpreted
cautiously.

The absence of a comparable $\gamma$-ray enhancement elsewhere may indicate either
a lower cosmic-ray density or a smaller covering fraction of dense gas in those
quadrants, and distinguishing these possibilities requires higher-resolution CO and
H$\alpha$ data. Such observations would locate individual clumps and constrain their
densities, and deeper GeV and TeV data would pin down the proton cutoff and the
hadronic contribution.

\section{Conclusion}
\label{sec:conclusion}

Using 17 years of Fermi-LAT data, complemented by radio and CO observations and by published X-ray constraints, we have 
detected and characterised extended $\gamma$-ray emission toward the middle-aged SNR 
G156.2+05.7. Our main results are as follows.

\begin{enumerate}
\item The $\gamma$-ray emission is spatially extended ($R_{68} \sim 0.84^\circ$) and shows evidence for curvature 
($\approx 3.4\sigma$ preference for a log-parabola over a power law). No long-term flux variability 
is detected, consistent with a diffuse origin.
\item Morphological tests favour a 2D Gaussian spatial model over the $0$--$10$ km s$^{-1}$ foreground CO template. This gas lies at kinematic distances of $0.5$--$1.0$ kpc and is therefore treated as a foreground line-of-sight structure rather than as the physical target of the emission.
\item For the adopted reference distance 
$d = 2.7$ kpc and under our one-zone assumptions, the joint SED fit gives the
leptonic--hadronic model the lowest BIC. It improves over the leptonic-only model by
$\Delta\mathrm{BIC} = 30.2$, but over the hadronic-plus-synchrotron model by only
$\Delta\mathrm{BIC} = 6.8$. Thus, while a hadronic contribution is favoured over a
purely leptonic interpretation, a proton-dominated origin remains competitive and the
partition of the GeV emission is model-dependent. The electron population provides the
radio synchrotron emission and a leptonic $\gamma$-ray component, while a steep proton population
with a cutoff fixed by assumption at $E_{p,\text{cut}} = 1$ TeV contributes the hadronic 
$\gamma$-rays.
\item The hadronic energy budget exceeds the canonical supernova energy
 for a single emission region at the tenuous density of the remnant interior
 ($n = 0.036$ cm$^{-3}$), even after allowing the maximal strong-shock
 compression of that gas, so the required conversion efficiency approaches
 or exceeds unity unless denser gas contributes to the emitting volume, well
 above the representative $\sim 10$ per cent acceleration-efficiency benchmark adopted
 in Section~\ref{sec:disc:scenarios} \citep{dermer2013gamma,zhang2019electron}
 (Fig.~\ref{fig:wp_vs_n}). These energy requirements are conditional on the adopted
 leptonic--hadronic decomposition, $E_{p,\mathrm{cut}}=1$ TeV, and
 $B=14\,\mu\mathrm{G}$; the proton-dominated alternative has a larger $A_{pp}$ and would
 make the energy budget more severe. For the adopted model, the electron cutoff ($\sim 1.2$ TeV) lies
 below the value that would produce detectable non-thermal X-ray synchrotron
 emission, so it is consistent with, though not constrained by, the
 thermal-dominated X-ray spectrum.
\item The two high-significance quadrants indicate internal structure. The northeast and southwest quadrants show $\gamma$-ray 
enhancement in directions that correspond in azimuth to the optical H$\alpha$ filaments. This asymmetry may reflect 
localised density or shock-environment inhomogeneities. This quadrant-level result is suggestive rather than
conclusive, because the quadrant definition, the diffuse background, and the limited
statistics of the fainter quadrants may affect the inferred asymmetry.
\end{enumerate}

G156.2+05.7 thus provides a case study of a middle-aged SNR whose $\gamma$-ray emission is described by a leptonic--hadronic model under the adopted assumptions. Its energy content is density-dependent. A single emission region at the tenuous interior density would require essentially the entire explosion energy, even after maximal compression. Therefore, unless the acceleration efficiency exceeds the representative $\sim 10$ per cent benchmark adopted in Section~\ref{sec:disc:scenarios}, the emitting volume would require denser gas than the tenuous interior, for example in a multiphase or clumpy medium. The SED alone does not determine the target density or its filling factor. Deeper TeV observations and improved target-density diagnostics will help constrain the particle acceleration and escape physics in this and similarly evolved remnants.

\section*{Acknowledgements}
We thank the anonymous referee for helpful comments and suggestions. This work is supported by the National Natural Science Foundation of China (Grant Nos. 12573045, 12563009 and 12303050)
and the Jiangxi Provincial Natural Science Foundation (Grant Nos. 20242BAB26012 and 20252BAC200593).
This research has made use of NASA's SkyView facility (https://skyview.gsfc.nasa.gov), located at NASA Goddard Space Flight Center, for the H$\alpha$ full-sky map data.

\section*{Data Availability}

The Fermi-LAT data underlying this article are publicly available via the NASA Fermi Science Support Center
at \url{https://fermi.gsfc.nasa.gov/ssc/}. The CO data are from the whole-Galaxy survey of \citet{dame2001milky}.
The H$\alpha$ map was retrieved from NASA's SkyView facility based on the full-sky composite of \citet{finkbeiner2003fullsky}.
The radio and Suzaku data were obtained from previously published sources cited in the text. The derived data products generated in this research---the posterior result files (\texttt{*\_results.ecsv}) and corner plots for all emission models and robustness checks---are provided as online supplementary material accompanying this article.

%%%%%%%%%%%%%%%%%%%%%%%%%%%%%%%%%%%%%%%%%%%%%%%%%%
%%%%%%%%%%%%%%%%%%%%%%%%%%%%%%%%%%%%%%%%%%%%%%%%%%

%%%%%%%%%%%%%%%%%%%% REFERENCES %%%%%%%%%%%%%%%%%%

\bibliographystyle{mnras}
\enlargethispage{2\baselineskip}
\bibliography{main}

\begin{thebibliography}{}
\makeatletter
\relax
\def\mn@urlcharsother{\let\do\@makeother \do\$\do\&\do\#\do\^\do\_\do\%\do\~}
\def\mn@doi{\begingroup\mn@urlcharsother \@ifnextchar [ {\mn@doi@}
  {\mn@doi@[]}}
\def\mn@doi@[#1]#2{\def\@tempa{#1}\ifx\@tempa\@empty \href
  {http://dx.doi.org/#2} {doi:#2}\else \href {http://dx.doi.org/#2} {#1}\fi
  \endgroup}
\def\mn@eprint#1#2{\mn@eprint@#1:#2::\@nil}
\def\mn@eprint@arXiv#1{\href {http://arxiv.org/abs/#1} {{\tt arXiv:#1}}}
\def\mn@eprint@dblp#1{\href {http://dblp.uni-trier.de/rec/bibtex/#1.xml}
  {dblp:#1}}
\def\mn@eprint@#1:#2:#3:#4\@nil{\def\@tempa {#1}\def\@tempb {#2}\def\@tempc
  {#3}\ifx \@tempc \@empty \let \@tempc \@tempb \let \@tempb \@tempa \fi \ifx
  \@tempb \@empty \def\@tempb {arXiv}\fi \@ifundefined
  {mn@eprint@\@tempb}{\@tempb:\@tempc}{\expandafter \expandafter \csname
  mn@eprint@\@tempb\endcsname \expandafter{\@tempc}}}

\bibitem[\protect\citeauthoryear{Abdollahi et~al.,}{Abdollahi
  et~al.}{2022}]{abdollahi2022incremental}
Abdollahi S.,  et~al., 2022, \mn@doi [The Astrophysical Journal Supplement
  Series] {10.3847/1538-4365/ac6751}, 260, 53

\bibitem[\protect\citeauthoryear{Ackermann et~al.,}{Ackermann
  et~al.}{2013}]{ackermann2013detection}
Ackermann M.,  et~al., 2013, \mn@doi [Science] {10.1126/science.1231160}, 339,
  807

\bibitem[\protect\citeauthoryear{Aharonian \& Atoyan}{Aharonian \&
  Atoyan}{1996}]{aharonian1996emissivity}
Aharonian F.~A.,  Atoyan A.~M.,  1996, Astronomy and Astrophysics, 309, 917

\bibitem[\protect\citeauthoryear{Akaike}{Akaike}{1974}]{akaike1974new}
Akaike H.,  1974, \mn@doi [IEEE Transactions on Automatic Control]
  {10.1109/TAC.1974.1100705}, 19, 716

\bibitem[\protect\citeauthoryear{Atwood et~al.,}{Atwood
  et~al.}{2009}]{atwood2009large}
Atwood W.~B.,  et~al., 2009, \mn@doi [The Astrophysical Journal]
  {10.1088/0004-637X/697/2/1071}, 697, 1071

\bibitem[\protect\citeauthoryear{Ballet, Bruel, Burnett, Lott  \& {The
  Fermi-LAT collaboration}}{Ballet et~al.}{2023}]{ballet2023fermi}
Ballet J.,  Bruel P.,  Burnett T.~H.,  Lott B.,   {The Fermi-LAT collaboration}
  2023, Fermi Large Area Telescope Fourth Source Catalog Data Release 4
  (4FGL-DR4), \mn@doi{10.48550/arXiv.2307.12546}

\bibitem[\protect\citeauthoryear{Bell}{Bell}{1978}]{bell1978acceleration}
Bell A.~R.,  1978, \mn@doi [Monthly Notices of the Royal Astronomical Society]
  {10.1093/mnras/182.2.147}, 182, 147

\bibitem[\protect\citeauthoryear{Bell}{Bell}{2004}]{bell2004turbulent}
Bell A.~R.,  2004, \mn@doi [Monthly Notices of the Royal Astronomical Society]
  {10.1111/j.1365-2966.2004.08097.x}, 353, 550

\bibitem[\protect\citeauthoryear{Blandford \& Eichler}{Blandford \&
  Eichler}{1987}]{blandford1987particle}
Blandford R.,  Eichler D.,  1987, \mn@doi [Physics Reports]
  {10.1016/0370-1573(87)90134-7}, 154, 1

\bibitem[\protect\citeauthoryear{Blasi}{Blasi}{2013}]{blasi2013origin}
Blasi P.,  2013, \mn@doi [The Astronomy and Astrophysics Review]
  {10.1007/s00159-013-0070-7}, 21, 70

\bibitem[\protect\citeauthoryear{Bolatto, Wolfire  \& Leroy}{Bolatto
  et~al.}{2013}]{bolatto2013cotoh2}
Bolatto A.~D.,  Wolfire M.,   Leroy A.~K.,  2013, \mn@doi [Annual Review of
  Astronomy and Astrophysics] {10.1146/annurev-astro-082812-140944}, 51, 207

\bibitem[\protect\citeauthoryear{Brand \& Blitz}{Brand \&
  Blitz}{1993}]{brand1993velocity}
Brand J.,  Blitz L.,  1993, Astronomy and Astrophysics, 275, 67

\bibitem[\protect\citeauthoryear{Cao et~al.,}{Cao et~al.}{2024}]{cao2024first}
Cao Z.,  et~al., 2024, \mn@doi [The Astrophysical Journal Supplement Series]
  {10.3847/1538-4365/acfd29}, 271, 25

\bibitem[\protect\citeauthoryear{Cardillo, Amato  \& Blasi}{Cardillo
  et~al.}{2016}]{cardillo2016supernova}
Cardillo M.,  Amato E.,   Blasi P.,  2016, \mn@doi [Astronomy and Astrophysics]
  {10.1051/0004-6361/201628669}, 595, A58

\bibitem[\protect\citeauthoryear{Dame, Hartmann  \& Thaddeus}{Dame
  et~al.}{2001}]{dame2001milky}
Dame T.~M.,  Hartmann D.,   Thaddeus P.,  2001, \mn@doi [The Astrophysical
  Journal] {10.1086/318388}, 547, 792

\bibitem[\protect\citeauthoryear{Dermer \& Powale}{Dermer \&
  Powale}{2013}]{dermer2013gamma}
Dermer C.~D.,  Powale G.,  2013, \mn@doi [Astronomy and Astrophysics]
  {10.1051/0004-6361/201220394}, 553, A34

\bibitem[\protect\citeauthoryear{Drury, Aharonian  \& Voelk}{Drury
  et~al.}{1994}]{drury1994gammaray}
Drury L.~O.,  Aharonian F.~A.,   Voelk H.~J.,  1994, The gamma-ray visibility
  of supernova remnants. A test of cosmic ray origin,
  \mn@doi{10.48550/arXiv.astro-ph/9305037}

\bibitem[\protect\citeauthoryear{Fesen \& Kirshner}{Fesen \&
  Kirshner}{1980}]{fesen1980spectrophotometry}
Fesen R.~A.,  Kirshner R.~P.,  1980, \mn@doi [The Astrophysical Journal]
  {10.1086/158534}, 242, 1023

\bibitem[\protect\citeauthoryear{Finkbeiner}{Finkbeiner}{2003}]{finkbeiner2003fullsky}
Finkbeiner D.~P.,  2003, \mn@doi [The Astrophysical Journal Supplement Series]
  {10.1086/374411}, 146, 407

\bibitem[\protect\citeauthoryear{{Foreman-Mackey}, Hogg, Lang  \&
  Goodman}{{Foreman-Mackey} et~al.}{2013}]{foreman-mackey2013emcee}
{Foreman-Mackey} D.,  Hogg D.~W.,  Lang D.,   Goodman J.,  2013, \mn@doi
  [Publications of the Astronomical Society of the Pacific] {10.1086/670067},
  125, 306

\bibitem[\protect\citeauthoryear{Funk}{Funk}{2015}]{funk2015ground}
Funk S.,  2015, \mn@doi [Annual Review of Nuclear and Particle Science]
  {10.1146/annurev-nucl-102014-022036}, 65, 245

\bibitem[\protect\citeauthoryear{Gabici, Aharonian  \& Casanova}{Gabici
  et~al.}{2009}]{gabici2009broadband}
Gabici S.,  Aharonian F.~A.,   Casanova S.,  2009, \mn@doi [Monthly Notices of
  the Royal Astronomical Society] {10.1111/j.1365-2966.2009.14832.x}, 396, 1629

\bibitem[\protect\citeauthoryear{Gerardy \& Fesen}{Gerardy \&
  Fesen}{2007}]{gerardy2007discovery}
Gerardy C.~L.,  Fesen R.~A.,  2007, \mn@doi [Monthly Notices of the Royal
  Astronomical Society] {10.1111/j.1365-2966.2007.11494.x}, 376, 929

\bibitem[\protect\citeauthoryear{Hartigan, Raymond  \& Hartmann}{Hartigan
  et~al.}{1987}]{hartigan1987radiative}
Hartigan P.,  Raymond J.,   Hartmann L.,  1987, \mn@doi [The Astrophysical
  Journal] {10.1086/165204}, 316, 323

\bibitem[\protect\citeauthoryear{Katsuda, Tanaka, Morokuma, Fesen  \&
  Milisavljevic}{Katsuda et~al.}{2016}]{katsuda2016constraining}
Katsuda S.,  Tanaka M.,  Morokuma T.,  Fesen R.,   Milisavljevic D.,  2016,
  \mn@doi [The Astrophysical Journal] {10.3847/0004-637X/826/2/108}, 826, 108

\bibitem[\protect\citeauthoryear{Kun, Kiss  \& Balog}{Kun
  et~al.}{2008}]{kun2008star}
Kun M.,  Kiss Z.~T.,   Balog Z.,  2008, Star Forming Regions in Cepheus,
  \mn@doi{10.48550/arXiv.0809.4761}

\bibitem[\protect\citeauthoryear{Li \& Chen}{Li \& Chen}{2010}]{li2010grays}
Li H.,  Chen Y.,  2010, \mn@doi [Monthly Notices of the Royal Astronomical
  Society] {10.1111/j.1745-3933.2010.00944.x}, 409, L35

\bibitem[\protect\citeauthoryear{Murase et~al.,}{Murase
  et~al.}{2026}]{murase2026gas}
Murase T.,  et~al., 2026, \mn@doi [The Astrophysical Journal]
  {10.3847/1538-4357/ae3dd4}, 1001, 19

\bibitem[\protect\citeauthoryear{Nolan et~al.,}{Nolan
  et~al.}{2012}]{nolan2012fermi}
Nolan P.~L.,  et~al., 2012, \mn@doi [The Astrophysical Journal Supplement
  Series] {10.1088/0067-0049/199/2/31}, 199, 31

\bibitem[\protect\citeauthoryear{Orlando, Bocchino, Reale, Peres  \&
  Petruk}{Orlando et~al.}{2007}]{orlando2007origin}
Orlando S.,  Bocchino F.,  Reale F.,  Peres G.,   Petruk O.,  2007, \mn@doi
  [Astronomy \& Astrophysics] {10.1051/0004-6361:20066045}, 470, 927

\bibitem[\protect\citeauthoryear{Pannuti \& Allen}{Pannuti \&
  Allen}{2004}]{pannuti2004asca}
Pannuti T.~G.,  Allen G.~E.,  2004, \mn@doi [Advances in Space Research]
  {10.1016/j.asr.2003.07.026}, 33, 434

\bibitem[\protect\citeauthoryear{Pfeffermann, Aschenbach  \&
  Predehl}{Pfeffermann et~al.}{1991}]{pfeffermann1991discovery}
Pfeffermann E.,  Aschenbach B.,   Predehl P.,  1991, Astronomy and
  Astrophysics, 246, L28

\bibitem[\protect\citeauthoryear{Porter, Moskalenko  \& Strong}{Porter
  et~al.}{2006}]{porter2006inverse}
Porter T.~A.,  Moskalenko I.~V.,   Strong A.~W.,  2006, \mn@doi [The
  Astrophysical Journal] {10.1086/507770}, 648, L29

\bibitem[\protect\citeauthoryear{Reich, Fuerst  \& Arnal}{Reich
  et~al.}{1992}]{reich1992radio}
Reich W.,  Fuerst E.,   Arnal E.~M.,  1992, Astronomy and Astrophysics, 256,
  214

\bibitem[\protect\citeauthoryear{Reid et~al.,}{Reid
  et~al.}{2014}]{reid2014trigonometric}
Reid M.~J.,  et~al., 2014, \mn@doi [The Astrophysical Journal]
  {10.1088/0004-637X/783/2/130}, 783, 130

\bibitem[\protect\citeauthoryear{Reynolds}{Reynolds}{2008}]{reynolds2008supernova}
Reynolds S.~P.,  2008, \mn@doi [Annual Review of Astronomy and Astrophysics]
  {10.1146/annurev.astro.46.060407.145237}, 46, 89

\bibitem[\protect\citeauthoryear{Uchida, Tsunemi, Katsuda, Mori, Petre  \&
  Yamaguchi}{Uchida et~al.}{2012}]{uchida2012suzaku}
Uchida H.,  Tsunemi H.,  Katsuda S.,  Mori K.,  Petre R.,   Yamaguchi H.,
  2012, \mn@doi [Publications of the Astronomical Society of Japan]
  {10.1093/pasj/64.3.61}, 64, 61

\bibitem[\protect\citeauthoryear{Uchiyama, Blandford, Funk, Tajima  \&
  Tanaka}{Uchiyama et~al.}{2010}]{uchiyama2010gammaray}
Uchiyama Y.,  Blandford R.~D.,  Funk S.,  Tajima H.,   Tanaka T.,  2010,
  \mn@doi [The Astrophysical Journal Letters] {10.1088/2041-8205/723/1/L122},
  723, L122

\bibitem[\protect\citeauthoryear{Wood, Caputo, Charles, Di~Mauro, Magill,
  Perkins  \& {Fermi-LAT Collaboration}}{Wood et~al.}{2017}]{wood2017fermipy}
Wood M.,  Caputo R.,  Charles E.,  Di~Mauro M.,  Magill J.,  Perkins J.~S.,
  {Fermi-LAT Collaboration} 2017, in 35th International Cosmic Ray Conference
  (ICRC2017). eprint: arXiv:1707.09551, p.~824, \mn@doi{10.22323/1.301.0824}

\bibitem[\protect\citeauthoryear{Xu, Han, Sun, Reich, Xiao, Reich  \&
  Wielebinski}{Xu et~al.}{2007}]{xu2007polarization}
Xu J.~W.,  Han J.~L.,  Sun X.~H.,  Reich W.,  Xiao L.,  Reich P.,   Wielebinski
  R.,  2007, \mn@doi [Astronomy \& Astrophysics] {10.1051/0004-6361:20077549},
  470, 969

\bibitem[\protect\citeauthoryear{Zabalza}{Zabalza}{2015}]{zabalza2015naima}
Zabalza V.,  2015, \mn@doi [Proc. of International Cosmic Ray Conference 2015]
  {10.48550/arXiv.1509.03319}, p.~922

\bibitem[\protect\citeauthoryear{Zhang \& Liu}{Zhang \&
  Liu}{2019}]{zhang2019electron}
Zhang X.,  Liu S.,  2019, \mn@doi [The Astrophysical Journal]
  {10.3847/1538-4357/ab14df}, 876, 24

\makeatother
\end{thebibliography}

% flushend column balancing conflicts with floats on the final page (a float can be dropped),
% so it is disabled from the appendix onwards where float-bearing pages end the paper.
\raggedend
\clearpage
\appendix
\section{Supplementary Robustness Checks}
\label{sec:app}

\subsection{Morphology checks}
\label{sec:app:morphology}

\begin{inplacetable}
\centering
\small
\caption{\label{tab:morphology_checks}Morphology robustness checks for the $\gamma$-ray
emission. PA denotes the position angle of the elliptical template.}
\begin{tabular}{@{}llll@{}}
\hline
Energy range & Template & Fitted morphology & $\Delta\ln\mathcal{L}$\textsuperscript{a}\\
\hline
100 MeV--1 TeV & Circular Gaussian & $R_{68}=0\fdg84^{+0.09}_{-0.09}$ & 0\\
100 MeV--1 TeV & \makecell[l]{Best elliptical\\template\textsuperscript{b}} &
\makecell[l]{$R_{68}=0\fdg95\times0\fdg65$;\\ PA $=44\fdg75$} & $\approx3$\\
$>1$ GeV & Circular Gaussian & $R_{68}=1\fdg05^{+0.13}_{-0.12}$ & --\\
\hline
\end{tabular}
\vspace{4pt}

\textsuperscript{a}Relative to the full-band circular Gaussian fit; the $>1$ GeV fit
uses a different energy range and is therefore not compared by likelihood to the full-band
models.

\textsuperscript{b}The better of the elliptical Gaussian and elliptical disc trials.
The likelihood gain corresponds to two additional shape parameters and is not decisive under
the AIC comparison.
\end{inplacetable}

\subsection{Magnetic field dependence}

\label{sec:app:bfield}

\begin{figure}
    \centering
    \includegraphics[width=\columnwidth]{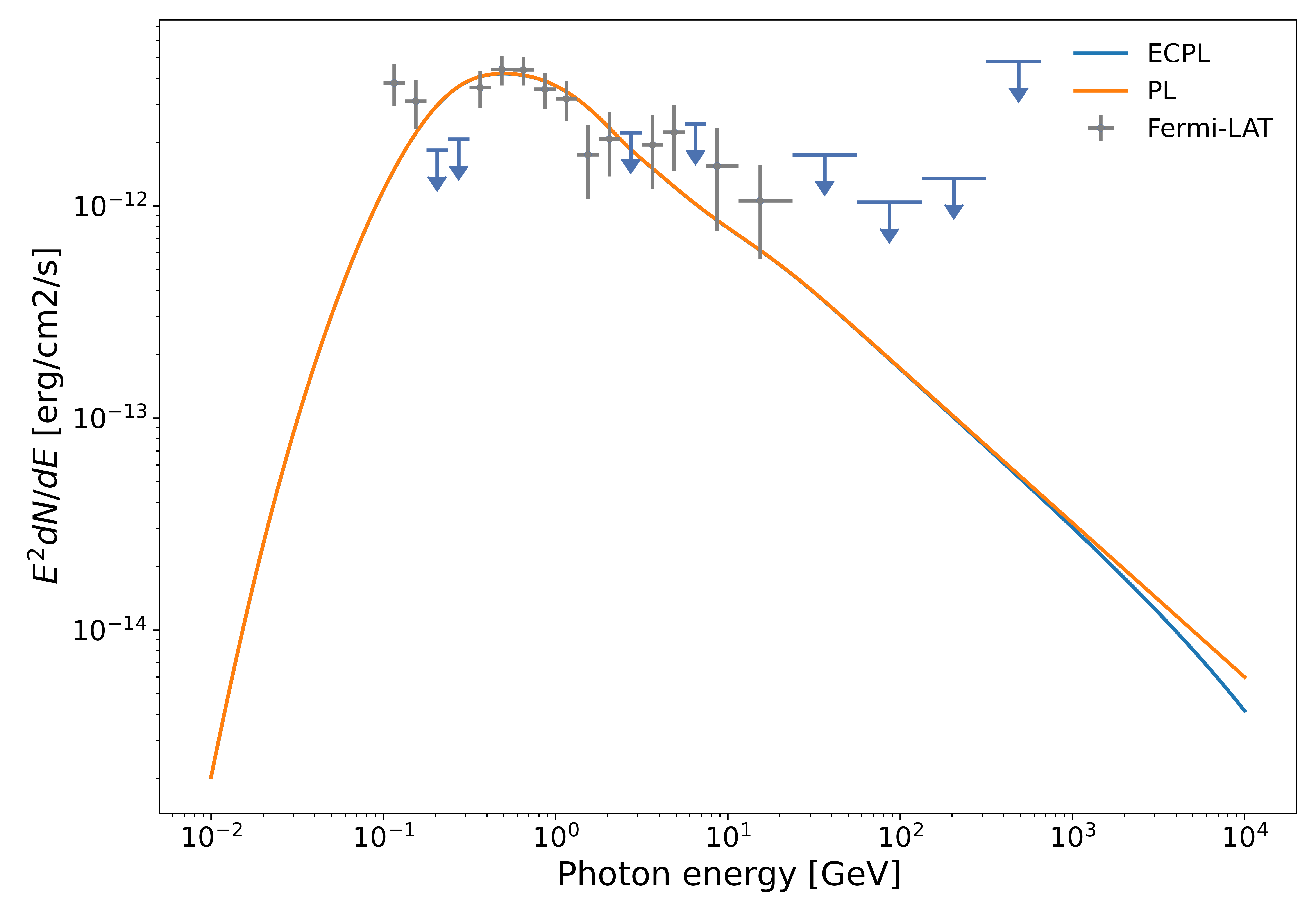}
    \caption{Pure hadronic (proton-only) emission fitted to the Fermi-LAT
    data alone, without radio or other multi-wavelength constraints. The solid
    and dashed curves show proton spectra with an exponential cutoff and a pure power
    law with no cutoff, respectively. The two curves are indistinguishable across the
    Fermi-LAT band, with only slight separation at the high-energy tail,
    illustrating that $E_{p,\mathrm{cut}}$ is not constrained by the Fermi-LAT data quality.}
    \label{fig:had_cutoff_compare}
\end{figure}

\begin{figure}
    \centering
    \includegraphics[width=\columnwidth]{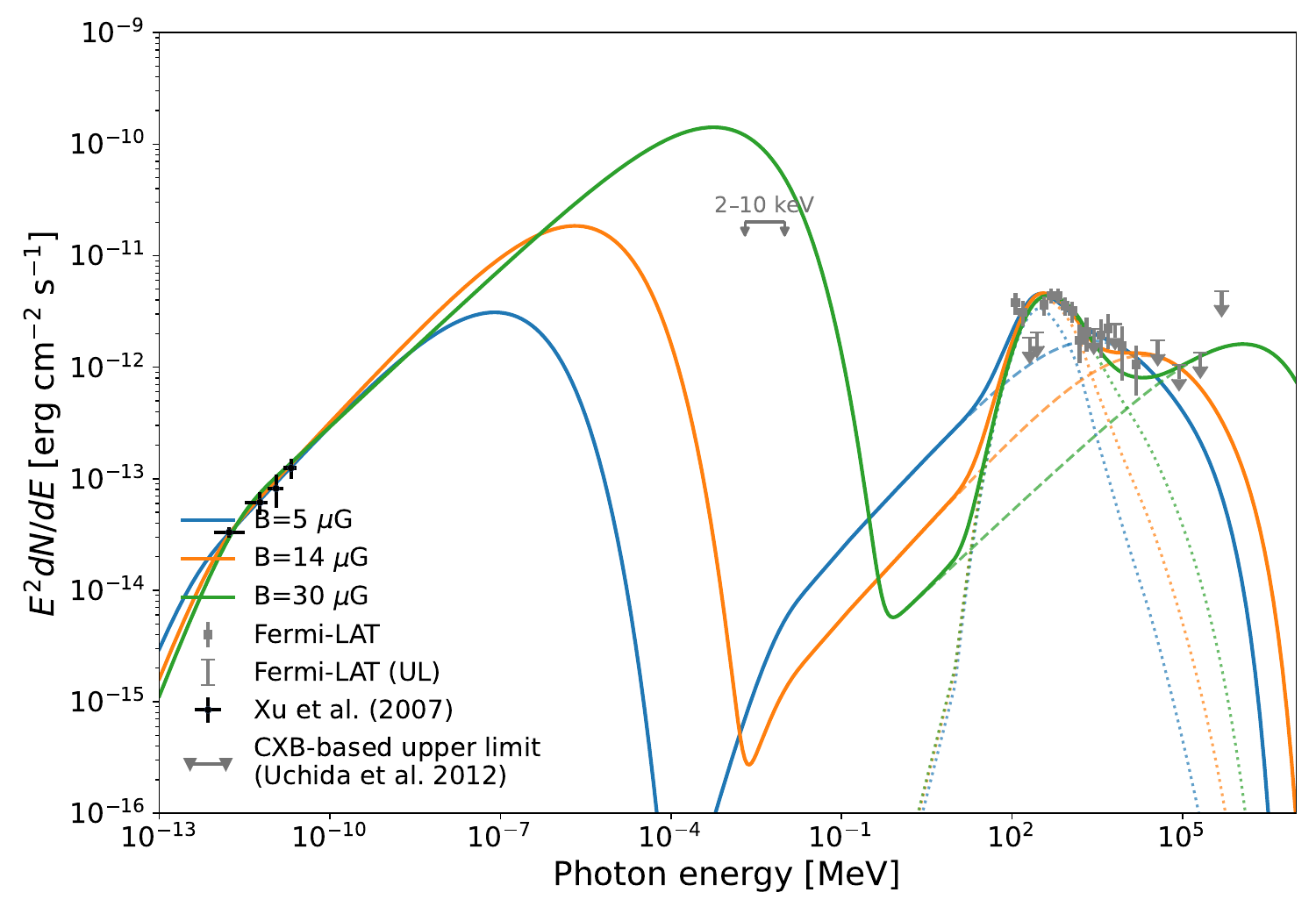}
    \caption{Leptonic--hadronic SED fits for $B = 5$, 14,
    and $30\,\mu\mathrm{G}$, used to test the magnetic-field dependence of the
    hadronic component. Solid, dashed, and dotted curves show the total, IC,
    and $\pi^{0}$-decay components, respectively. The radio measurements entering
    the fits are shown; the same data also appear in the joint-fit panels of
    Fig.~\ref{fig:combined_fit_and_plot}.}
    \label{fig:bfield_sed_comparison}
\end{figure}

We repeat the leptonic--hadronic joint fit for $B = 5$, $14$, and
$30\,\mu\mathrm{G}$, holding $B$ fixed and $E_{p,\mathrm{cut}} = 1$ TeV as in
the baseline analysis (Section~\ref{sec:disc:sed};
Fig.~\ref{fig:had_cutoff_compare}). The BIC, evaluated at the maximum-likelihood
point, decreases toward weaker fields ($\Delta\mathrm{BIC}\approx3$ between $B=5$
and $14\,\mu\mathrm{G}$, and $\approx9$ between the extremes;
Table~\ref{tab:bfield_variation}); because the $B=5$ and $B=14$ fits are not
significantly separated, the reference value $B=14\,\mu\mathrm{G}$ is kept as the
baseline. The hadronic component persists at all three field strengths
(Fig.~\ref{fig:bfield_sed_comparison}). The energy-budget
argument of Section~\ref{sec:disc:scenarios} is likewise insensitive to $B$:
$W_p(>1.22\,\mathrm{GeV})$ at $n = 1$ cm$^{-3}$ is $1.37$, $1.45$, and
$1.35\times10^{50}$ erg for $B = 5$, $14$, and $30\,\mu\mathrm{G}$
($\varepsilon_{\rm cr} \simeq 0.14\,(n/1\,\mathrm{cm^{-3}})^{-1}$, a $\sim7$ per
cent spread), because weaker-field fits pair a smaller $A_{pp}$ with a steeper
$\alpha_p$, keeping the proton energy fixed where the Fermi-LAT SED anchors the
spectrum. The tenuous single-zone case ($n = 0.036$ cm$^{-3}$) still requires
$\approx 3.7$--$4.0\,E_{\rm SN}$, so denser target gas remains necessary at all
$B$. The electron energy content, by contrast, falls from $1.9\times10^{48}$ erg
at $B = 5\,\mu\mathrm{G}$ to $9.8\times10^{46}$ erg at $B = 30\,\mu\mathrm{G}$,
as expected from the stronger synchrotron emissivity at higher fields.

\begin{inplacetable}
\centering
\small
\caption{\label{tab:bfield_variation}Magnetic-field dependence of the
leptonic--hadronic joint fit for $B = 5$, $14$, and $30\,\mu\mathrm{G}$.}
\begin{tabular}{@{}lr@{}}
\hline
$B$ ($\mu$G) & BIC\textsuperscript{a}\\
\hline
5  & 35.58\\
14 & 38.62\\
30 & 44.59\\
\hline
\end{tabular}
\vspace{4pt}

\textsuperscript{a}All fits use the Fermi-LAT+radio dataset; the BIC values are computed from the maximum-likelihood fit of each fixed-$B$ model and do not constitute a Bayesian evidence ratio (see text).
\end{inplacetable}

%%%%%%%%%%%%%%%%%%%%%%%%%%%%%%%%%%%%%%%%%%%%%%%%%%

\bsp
\label{lastpage}
\end{document}